\documentclass[12pt]{article}
\pdfoutput=1
\usepackage{paper12}
\usepackage{comment}
\usepackage{setspace}
\usepackage[normalem]{ulem} 

\usepackage{caption}
\newcommand{\Emax}{E_\text{max}}
\newcommand{\oV}{\op{V}}

\renewcommand{\D}{\mathrm{d}}

\begin{document}

\begin{titlepage}
{\raggedleft \vphantom{x}\\[-30pt] CERN-TH-2026-293\\ IPPP/26/67\par}
\title{Chiral Gauge Theory on a Computer}

\author{Timothy Cohen}

\address{
Institute for Fundamental Science\\[-2pt]
University of Oregon, Eugene, Oregon 97403, USA\\[2pt]
Theoretical Physics Department, CERN, 1211 Geneva, CH
}

\author{Kara Farnsworth}

\address{Maxwell Institute for Mathematical Sciences, Department of Mathematics\\[-2pt]
Heriot-Watt University, Edinburgh EH14, UK}

\author{Rachel Houtz}

\address{Department of Physics, University of Florida, Gainesville, FL 32611, USA}

\author{James Ingoldby}

\address{Institute for Particle Physics Phenomenology \\[-2pt]
Durham University, Durham DH1 3LE, UK \\[2pt]
Theoretical Physics Department, CERN, 1211 Geneva, CH}

\author{Markus A.~Luty}

\address{Center for Quantum Mathematics and Physics (QMAP)\\[-2pt]
University of California, Davis, California 95616, USA}

\begin{abstract}
We give a completely gauge-fixed
Hamiltonian formulation of a general anomaly-free 1+1 dimensional $U(1)$ chiral gauge theory on $S^1 \times \mathbb{R}$ and use it to obtain numerical results for the spectrum of the Schwinger model and the  `345 model' using Hamiltonian truncation. 
The Hamiltonian describes fermions interacting through a Coulomb term and coupled to a dynamical gauge zero mode. 
We give an explicit description of the physical states and formulas for the matrix elements of the Hamiltonian and other gauge invariant operators. The gauge anomalies of the theory appear as consistency conditions on the formulation,
and the 't Hooft anomalies are reproduced. Our construction makes clear that $\bar\th$ is a parameter of the quantum theory rather than a state-dependent quantity. The numerical results obtained in the fermionic description (obtained on a laptop) agree accurately with the bosonized solution, which we review.
\end{abstract}

\end{titlepage}

\noindent
\setcounter{page}{2}
\setcounter{tocdepth}{2}
\begin{spacing}{1.3}
\tableofcontents
\end{spacing}
\clearpage

\section{Introduction}
\label{sec:intro}
Developing a non-perturbative formulation of chiral gauge theories
suitable for numerical calculations remains an important
problem in theoretical physics.  The Standard Model is itself a
chiral gauge theory, and understanding the strongly coupled dynamics
of more general chiral theories could reveal novel phenomena relevant to
physics beyond the Standard Model and/or inform formal investigations.  The inherent difficulty is often 
described as finding a method for regulating the UV that respects both gauge
invariance and chiral symmetry.  These requirements can clash 
due to gauge anomalies, and implementing them consistently
is challenging even when the gauge anomalies cancel.

Much of the work on the non-perturbative formulation of chiral gauge theories has focused on lattice regularization.
Under its standard assumptions, the Nielsen--Ninomiya theorem~\cite{Nielsen:1980rz,Nielsen:1981xu} obstructs a local lattice realization of a chiral fermion spectrum; see Ref.~\cite{Golterman:2000hr} for a review. Approaches to this problem include realizing chiral fermions on defects or boundaries of a higher-dimensional theory~\cite{Callan:1984sa,Kaplan:1992bt,Shamir:1993zy,Grabowska:2015qpk,Kaplan:2023pxd,Kaplan:2023pvd,Kaplan:2024ezz,Sen:2026pnm}, and modifying the realization of chiral symmetry at finite lattice spacing through the Ginsparg--Wilson relation, as in the overlap formulation~\cite{Ginsparg:1981bj,Narayanan:1994gw,Neuberger:1997fp,Neuberger:1998wv,Luscher:1998pqa,Luscher:1999un,Grabowska:2016bis}. Building on these developments, Lüscher constructed a local, gauge invariant fermion measure non-perturbatively for anomaly-free Abelian theories and extended the framework to non-Abelian theories to all orders in perturbation theory~\cite{Luscher:1998du,Luscher:2000zd}. Recently, Craig proposed a non-perturbative extension of this construction to a class of non-Abelian chiral gauge theories~\cite{Craig:2026soq}. Related work uses continuum Weyl fermions coupled to lattice gauge fields~\cite{Hsu:2026gmi,Witten:2026ezr}, building on an earlier proposal of 't Hooft~\cite{tHooft:1994jaf}. Gauge-fixed approaches~\cite{Borrelli:1989kw,Bock:1997vc}, and the decoupling of mirror fermions through strong interactions that preserve the gauge symmetry~\cite{Eichten:1985ft}, have also been pursued.

Even when a consistent Euclidean formulation is available, the generally complex fermion weight poses a separate obstacle to efficient importance-sampling Monte Carlo calculations~\cite{Golterman:2000hr}.\footnote{Numerical calculations using the overlap formulation have been performed in a specific two-dimensional chiral gauge theory~\cite{NarayananNeuberger1997,KikukawaNarayananNeuberger1998}, where results are obtainable despite the complex-weight problem.
} This motivates computational approaches that do not rely on Euclidean importance sampling, including Hamiltonian methods. These include Hamiltonian lattice formulations~\cite{Quinn:1986wk,Wen:2013ppa,Wang:2018ugf,Thorngren:2026ydw,Seifnashri:2026ema,LuShao2026,Grabowska:2026zax} and tensor-network simulations of a 1+1 dimensional Abelian chiral gauge theory~\cite{Lootens:2026kcg}.

Here we write the Hamiltonian for the theory in the continuum and in finite volume, 
and use Hamiltonian truncation to compute non-perturbative observables in Abelian chiral gauge theories. 
In this approach it is straightforward to introduce fermions with exact chiral symmetry.
The price for this is that gauge invariance is not manifest;
in a way, this is the mirror of the difficulty on the lattice. We formulate the theory in terms of gauge invariant states. We do this by fixing the gauge, which amounts to choosing a representative for each physical state. The remainder of this introduction gives an overview of our methods.

\subsection{Overview}
The starting point is the Hamiltonian formulation of $U(1)$ chiral gauge
theories in 1+1 dimensions on a spatial circle $S^1$.
These theories have many features in common with non-Abelian gauge theories in 3+1 dimensions.
They are strongly coupled in the infrared, and they have
a nontrivial CP phase.
These theories exhibit interesting non-perturbative phenomena such as massive and massless composite states.
Our treatment of these theories largely follows the classic Hamiltonian treatments of the Schwinger model on $S^1$~\cite{Manton:1985jm,Hetrick:1988yg,Iso:1988zi}
and Hamiltonian studies of chiral
theories~\cite{Halliday:1985tg,GrosseLangmann2000}.
We work on $S^1$ rather than an interval because 
this is straightforwardly compatible with both gauge invariance and chiral symmetry.\footnote{Formulating chiral theories with boundaries has also been developed, see~\cite{Smith:2019jnh}.}

The theory is formulated in Coulomb gauge, which removes all gauge field degrees of freedom except a single zero mode parameterizing the holonomy around $S^1$.
Gauss' law is imposed by restricting to states with vanishing gauge charge and including a nonlocal Coulomb interaction.
The resulting Hamiltonian describes chiral fermions interacting with the gauge zero mode and via the Coulomb interaction.
We construct an explicit Fock space basis for the fermions and show how to compute matrix elements of the Hamiltonian in this basis.

This theory has a discrete residual gauge redundancy that we call large residual gauge transformations (LRGTs).
The LRGTs shift the fermion occupation numbers, yet they are gauge redundancies.
These transformations must be handled carefully; 
they are the source of several important subtleties of this theory.

Anomalies arise from the requirement that UV divergences must be regulated and renormalized in a way that is compatible with LRGTs.
Normal ordering the fermion operators removes UV divergences, but the resulting operators are not gauge invariant.
`Dressing' the conserved charges of the theory to make them gauge invariant can clash with conservation laws. 
This is the origin of the anomalies in this formulation.
We show that we can recover the standard gauge and ABJ anomalies in this way.

Next we turn to the construction of the physical states of the theory.
Physical states must be invariant under LRGTs, but the only invariant state of the theory described above is the zero state.
We instead define the physical states as LRGT equivalence classes of Coulomb-gauge states.
We show how this defines a Hilbert space of fully gauge invariant states with a well-defined positive-definite inner product.
We obtain simple explicit formulas for matrix elements of gauge invariant operators in physical states by choosing a representative for each physical state in the original Hilbert space.
These formulas give a completely gauge invariant formulation of the theory that is the starting point for numerical work.
In particular, we emphasize that truncating the physical space of states is completely gauge invariant.

The LRGTs play an important role in computing these matrix elements.
Gauge invariant operators that explicitly break chiral symmetry (for example a fermion mass term) give states outside of the representative space, and require a compensating LRGT to return them to the representative space.
This leads to the expected dependence on the physical CP-violating phase $\bar\th$.
It also clarifies why $\bar\th$ is a parameter of the theory rather than a property of physical states.

This class of theories was solved in the 1980s using bosonization~\cite{Halliday:1985tg,KutasovSchwimmer1995,GrosseLangmann2000}.
We give a self-contained discussion of this using our approach.
Bosonization
provides a useful comparison for numerical results, but we emphasize that our numerical results are obtained using the fermionic formulation of the theory.

We then perform numerical calculations of 
the interacting spectrum of the massless Schwinger model and chiral `345' model using Hamiltonian truncation~\cite{FitzpatrickKatz2022}. In this method, the Hamiltonian is written as a sum of a free Hamiltonian and an interacting Hamiltonian, and the eigenstates of the free Hamiltonian are used as a basis for the states of the interacting 
theory. By truncating this space, we approximate the eigenstates
of the full Hamiltonian. Because we are working with gauge invariant states, this truncation is gauge invariant. However, this formulation is neither manifestly Lorentz invariant
nor manifestly local (due to the nonlocal Coulomb interaction). 
However, we show that our numerical results agree very well with the bosonized spectrum.
The numerical convergence of energy level differences is compatible with $1/E_\mathrm{max}^3$, where $E_\mathrm{max}$ is the truncation cutoff.
This is the convergence predicted by the effective Hamiltonian arguments of \cite{Cohen:2021erm}, as we discuss below.

The simplicity of the two-dimensional theories
studied here makes them a
useful setting for developing this approach.  Extending the
construction to non-Abelian theories and higher dimensions 
is nontrivial,
particularly in gauge fixing and the treatment of UV divergences.  
These theories cannot be solved using bosonization, and thus verifying the Lorentz invariance and locality of the formulation will have to be done without recourse to an exact solution.
We emphasize that a completely gauge-fixed Hamiltonian formulation of chiral gauge theories may be a useful starting point for other numerical or analytical approaches to the theory.
The work reported here provides a starting point for these future investigations.

\section{Chiral $U(1)$ Gauge Theories on $S^1$}
We begin by defining the 1+1 dimensional Abelian chiral gauge theories we are studying in this paper. We compactify on $S^1$, where (anti)periodic boundary conditions are automatically compatible with gauge invariance and chiral symmetry of fermions.
We assume that the fermions are antiperiodic to eliminate their would-be zero modes.
The gauge transformations are periodic:
\beq
\Om(t, x + L) = \Om(t, x),
\eeq
where $L = 2\pi R$ is the circumference of the $S^1$.
The complex Weyl fermion $\psi^a$ with $a=1,\ldots,N$ has chirality $s_a=+1$
for a right mover and $s_a=-1$ for a left mover, and gauge charge $q_a\in\ZZ$.
(Our conventions are listed in the Appendix.)
Under a gauge transformation, we have
\bseq
\beq
\psi^a &\mapsto \Om^{q_a} \psi^a,
\\[4pt]
A_\mu &\mapsto A_\mu -\ii \d_\mu \log\Om.
\eeq
\eseq
The gauge covariant derivative is
\beq
D_\mu \psi^a = (\d_\mu - \ii q_a A_\mu) \psi^a.
\eeq
The classical Lagrangian density is 
\beq
\scr{L} &= -\frac{1}{4g^2} F^{\mu\nu} F_{\mu\nu}
+ \frac{\th}{4\pi} \ep^{\mu\nu} F_{\mu\nu}
+ \sum_a \psi^{a\dagger}\ii D_{s_a}\psi^a,
\eeq
where $g$ is the gauge coupling with mass dimension 1,
and $D_\pm = D_0 \pm D_1$ are the lightcone components of the gauge
covariant derivative ($D_s = D_0 + s D_1$).
We will use units where 
$R = 1$ so that $L = 2\pi$.

\subsection{Gauge Fixing}
We will define the quantum theory by
quantizing the gauge-fixed classical theory. We therefore turn next to the classical gauge fixing. The temporal component $A_0$ is an auxiliary field. We impose Coulomb gauge
\beq
\d_1 A_1 = 0.
\label{eq:Cgauge}
\eeq
With this choice, the only remaining dynamical gauge degree of freedom is the zero mode, which only depends on time:
\beq
A(t) \equiv A_1(t,x).
\eeq

The gauge-fixed Lagrangian density  is
\beq
\scr{L} &= \frac{1}{2g^2} \Big[
(\d_0 A)^2 + (\d_1 A_0)^2 \Big]
+ \frac{\th}{2\pi} \d_0 A
+ A_0 J^0 + A J^1 
+ \sum_a \psi^{a\dagger}\ii\d_{s_a}\psi^a,
\label{eq:LA0}
\eeq
where $\d_\pm = \d_0 \pm \d_1$. Here $J^\mu$ is the gauge current, given by
\beq
J^0(t,x) = \sum_a  q_a \rho^a(t,x)
\qquad\text{and}\qquad
J^1(t,x) = \sum_a  s_a\gap q_a \rho^a(t,x),
\label{eq:Jgauge}
\eeq
where
\beq
\label{eq:fcl}
\rho^a(t, x) \equiv (\psi^{a\dagger} \psi^a)(t,x)
\eeq
is the fermion number density for species $a$.
In writing \Eq{LA0}, we discarded total spatial derivatives
from both the Maxwell term and the $\th$ term, which is
justified on $S^1$.
The total time derivative proportional to $\th$ has been
kept.
Below we will show that it contributes to the physical theta
parameter $\bar\th$ in the quantized theory.

The gauge condition \Eq{Cgauge} is preserved by
\beq
\Om(t,x)=e^{\ii\al(t)}e^{\ii\nu x}
\qquad\text{with}\qquad
\nu\in\ZZ.
\eeq
The spatially constant transformations generated by $\al(t)$
shift the zero mode of $A_0$ and give the global Gauss law
constraint.
The remaining transformations
\beq
\Om_\nu(t,x)=e^{\ii\nu x}
\eql{LRGTs}
\eeq
form a subgroup of the infinite-dimensional group
of all gauge transformations that is
isomorphic to $\mathbb{Z}$, since
\beq
\Om_\nu(t,x)\gap \Om_{\nu'}(t,x) =\Om_{\nu+\nu'}(t,x).
\eeq
We refer to 
the transformations in \Eq{LRGTs}
as large residual gauge transformations (LRGTs),
because they preserve \Eq{Cgauge} and cannot be deformed to 
the identity for $\nu\ne0$.

Under the LRGTs, $A_0$ is invariant and
\bseq
\beq
A(t) &\mapsto A(t) + \nu,
\\[4pt]
\psi^a(t,x) &\mapsto e^{\ii x \nu q_a} \psi^a(t,x).
\eeq
\eseq
We are in finite volume, 
so these gauge transformations are unambiguously
a redundancy of the theory.

We now integrate out $A_0$ in the classical theory. We work at fixed time $t$, so from now on we will drop the explicit $t$ dependence.
The zero mode of $A_0$ gives the constraint
\beq
Q = \int \! \D x\, J^0 = 0,
\eeq
while the nonzero modes generate a nonlocal Coulomb interaction term
in the Lagrangian (\emph{not} the Lagrangian density)
\beq
L_\mathrm{Coul} = -\frac{g^2}{2} \int \! \D x \, \D y\,
J^0(x) K(x-y) J^0(y),
\label{eq:LCoul}
\eeq
where
\beq
K(x) = \frac{1}{\pi} \sum_{k = 1}^\infty \frac{\cos(k x)}{k^2}
\eeq
satisfies
\beq
-\d_1^2 K(x) = \de(x) - \frac{1}{2\pi}.
\eeq
Therefore, $K(x)$ is the Green's function for charge configurations satisfying $Q = 0$.

We can write this theory in terms of fermion Fourier modes defined by
\beq
\psi^a_\ell
\equiv \frac{1}{\sqrt{2\pi}} \int_0^{2\pi} \D x\,
e^{-\ii \ell x}\, \psi^a(x),
\eeq
where $\ell = \pm \frac 12, \pm \frac 32, \ldots\,$ 
due to
our choice of antiperiodic boundary conditions.
We define
\bseq
\beq
\rho^a_k &\equiv \int \D x\, e^{-\ii k x}\, \rho^a(x)
= \sum_\ell \psi^{a\dagger}_\ell \psi^a_{\ell + k},
\label{eq:rhokcl}
\\[3pt]
J^0_k &\equiv \int \D x\, e^{-\ii k x}\, J^0(x)
= \sum_a q_a \rho^a_k,
\label{eq:J0kcl}
\eeq
\eseq
where $k = 0, \pm 1, \pm 2, \ldots\,$.
The normalization of the Fourier transforms is chosen so that
\beq
Q = J^0_{k=0} = \sum_a q_a \rho^a_{k=0}.
\eeq

The classical Hamiltonian in Coulomb gauge is then
\beq
H = H_\text{gauge} + H_\text{ferm} + H_\text{Coul},
\eeq
where 
\bseq
\label{eq:Hclass}
\beq
H_\mathrm{gauge}
&= \frac{g^2}{4\pi} \left(
\Pi - \th \right)^2,
\label{eq:gaugekinetic}
\\[6pt]
H_\mathrm{ferm} 
&=  \sum_{a,\ell}
s_a \big( \ell - q_a A \big) 
\psi^{a\dagger}_\ell \psi^a_\ell,
\label{eq:Hferm}
\\[2pt]
H_\mathrm{Coul}
&=  \frac{g^2}{4\pi} \sum_{k \,\ne\, 0}  \frac{J^0_{-k} J^0_k}{k^2},
\label{eq:HCoulC}
\eeq
\eseq
and
\beq
\Pi = \frac{\d L}{\d \dot{A}} = \frac{2\pi}{g^2} \dot{A} + \th
\eeq
is the momentum conjugate to $A$.
Note that $H_\text{gauge}$, $H_\text{ferm}$, and $H_\text{Coul}$
are separately invariant under LRGTs.

The classical Hamiltonian \Eqs{Hclass} generates the correct classical 
equations of motion subject to the gauge condition $\d_1 A_1 = 0$ and the constraint $Q = 0$.
It is the starting point for our quantization of the theory.

\subsection{Quantization}
\label{sec:Quant}
We now quantize the classical theory introduced in the last section. The gauge-fixed classical theory is written in terms of the gauge zero mode $A$ and the fermion modes
$\psi^a_\ell$.
The operators $\oA$ and $\oPi$ act
on complex wavefunctions $\Psi(A)$ in the usual way:
\beq
\oA \Psi(A) = A \Psi(A)
\qquad\text{and}\qquad
\op{\Pi} \Psi(A) = -\ii \frac{\d}{\d A} \Psi(A).
\eeq
The fermion fields are defined by their action on free-fermion Fock space.
The momentum-space fermion field operators satisfy the canonical anticommutation relations
\beq
\big\{ \opsi^{a\vphantom{\dagger}}_{\ell{\vphantom{'}}}, \opsi^{b\dagger}_{\ell'} \big\}
= \de^{ab} \de_{\ell\ell'}
\qquad\text{and}\qquad
\big\{ \opsi^a_{\ell{\vphantom{'}}}, \opsi^{b}_{\ell'} \big\} = 0.
\eeq
At $A = 0$, the energy of a single-particle fermion mode 
with momentum $\ell$ is $s_a\ell$, which can be arbitrarily negative.
This tells us that the ground state of the free theory can be obtained by filling the Dirac sea.
In practice, this implies that the fermion annihilation operator is defined as
\beq
\oa^a_\ell \equiv \begin{cases}
\opsi^a_\ell & s_a\ell > 0, \\[3pt]
\opsi^{a\dagger}_\ell & s_a\ell < 0.
\end{cases}
\eeq

We conclude that the space of free-particle states can be expressed as $\ket{A, n}$.
Here $A$ is the eigenvalue of $\oA$, 
\beq
\oA \ket{A, n} &= A \ket{A, n},
\eeq
while $n$ denotes the collection of
fermion occupation numbers $n^a_\ell = 0,1$
defined relative to the Dirac vacuum. A general Fock basis state is then given by
\beq
\ket{A, n} &= \prod_{a,\ell}
\big[ (\oa^a_\ell)^\dagger \big]^{n^a_\ell} \ket{A, 0},
\label{eq:Fockstate}
\eeq
where the Dirac vacuum $\ket{A, 0}$ satisfies 
\beq
\oa^a_\ell \ket{A, 0} = 0
\qquad
\text{for all $a$, $\ell$}.
\eeq

The LRGTs act on the states constructed above by a unitary transformation
$\oR_\nu$.
Its defining property is that it acts on the fields by
\bseq
\label{eq:LRGTfields}
\beq
\oR_\nu  \oA \oR_{-\nu}
&= \oA + \nu,
\label{eq:ARLRGT}
\\[3pt]
\oR_\nu  \opsi^a_\ell  \oR_{-\nu}
&= \opsi^a_{\ell - \nu q_a}.
\eeq
\eseq
This preserves the (anti)commutation relations, and defines a unitary action on
the states given by
\beq
\oR_\nu \ket{A, n}
= \eta_\nu(n)\ket{A - \nu, R_\nu(n)},
\label{eq:Rnudefn}
\eeq
where $\eta_\nu(n)$ is a phase and
$R_\nu(n)$ implements the LRGT on the fermion occupation
numbers:
\beq
R_\nu(n)^a_{\ell} = \begin{cases}
n^a_{\ell + \nu q_a} & \mathrm{sgn}(\ell + \nu q_a)
= \mathrm{sgn}(\ell), \\[3pt]
1 - n^a_{\ell + \nu q_a} & \text{otherwise}.
\end{cases}
\label{eq:Rn}
\eeq
Because the theory is defined relative to the Dirac sea,
the LRGT can turn an occupied mode into an unoccupied mode,
or {\it vice versa\/}.
The phases $\eta_\nu(n)$ depend on the ordering convention
for the fermion basis states; we will not need 
explicit expressions for these, because
we will use the relations \Eqs{LRGTfields}
to derive our results. We conclude that the states $\ket{A,n}$ are not invariant under LRGTs, so they are not
physical states. 
We will obtain the physical Hilbert space by choosing a representative state for each set of states related by LRGTs.

We define composite fermion operators on this Fock space by normal
ordering with respect to the Dirac vacuum.  The notation $\no{\osO}$
means that the creation operators $(\oa^a_\ell)^\dagger$ are placed
to the left of the annihilation operators $\oa^a_\ell$, along with the standard minus signs due to Fermi statistics.
In particular, we define the local
operator
\beq
\orho^a(x) &\equiv
\no{\opsi^{a\dagger}(x)\opsi^a(x)},
\eeq
and the integrated `charges'
\bseq
\beq
\orho^a_k &\equiv \int \D x\,e^{-\ii kx}\orho^a(x)
=\sum_\ell\no{\opsi^{a\dagger}_\ell\opsi^a_{\ell+k}},
\label{eq:rhokQ}
\\[3pt]
\of^a &\equiv \orho^a_{k=0},
\label{eq:fhat}
\\[6pt]
\oh^a &\equiv s_a\sum_\ell(\ell-q_a\oA)
\no{\opsi^{a\dagger}_\ell\opsi^a_\ell}.
\label{eq:fhNO}
\eeq
\eseq
Here $\of^a$ is the fermion number for species $a$, and $\oh^a$
is the single-particle Hamiltonian 
for fermion species $a$
coupled to $\oA$.  We also define
the normal ordered gauge currents by
\beq
\oJ^0(x)=\sum_a q_a\orho^a(x)
\qquad\text{and}\qquad
\oJ^1(x)=\sum_a s_a q_a\orho^a(x).
\eeq

Normal ordered products of the fermion fields and their derivatives
give a complete basis of local fermion operators.  They have finite
matrix elements between Fock states with a finite number of
excitations, and provide the operators used in our quantization.
Their equal-time (anti)commutators with elementary fermion fields
reproduce the normal ordered Poisson brackets exactly.
Commutators of two composite operators can nevertheless have
additional terms, as we will see in \sec{anomaly}.
Microcausality also requires Lorentz covariance of the operators
and Lorentz invariance of physical observables.  These properties
are not manifest in Coulomb gauge; in the case studied here,
we will check that the resulting quantum theory is correct by comparing against the exact solution derived using bosonization in \sec{Bosonization}.

\subsection{Covariantized Operators}
\label{sec:Regularization}
The normal ordered operators constructed above are not
necessarily invariant under LRGTs, even though the corresponding
classical operators are invariant.
The reason is that normal ordering is defined relative to the
Dirac vacuum, which is not invariant under these transformations.
In this subsection, we will compute the transformation law
of these operators under LRGTs and show how to `dress'
them to make them invariant under LRGTs.  
This construction does not require a UV regulator.

To see this explicitly, consider the diagonal fermion bilinear in
\Eq{fhNO}.  Applying \Eqs{LRGTfields} and restoring normal order
gives
\beq
\oR_\nu\no{\opsi^{a\dagger}_\ell\opsi^a_\ell}\oR_{-\nu}
&=\no{\opsi^{a\dagger}_{\ell-\nu q_a}
\opsi^a_{\ell-\nu q_a}}
+\frac{s_a}{2}
\big[\mathrm{sgn}(\ell)-\mathrm{sgn}(\ell-\nu q_a)\big].
\label{eq:LRGTbilinear}
\eeq
The last term is nonzero only for the $|\nu q_a|$ modes that cross
the Dirac sea.  
It is obtained using only a finite number of
nonzero anticommutators, and is therefore unambiguous without
any UV regulator.  
Summing these terms gives the
transformation of the density modes:
\beq
\oR_\nu\orho^a_k\oR_{-\nu}
=\orho^a_k+\nu s_aq_a\de_{k,0}.
\label{eq:LRGTrho}
\eeq
For $k\ne0$, reordering produces no anticommutator term, so
these modes are already invariant under LRGTs.

For $\oh^a$, the simultaneous shifts of $\ell$ and $\oA$ leave
the coefficient of the normal ordered bilinear unchanged after
relabeling the sum.  The only additional contribution is again
from the modes crossing the sea.  For $\nu q_a>0$, the affected
modes have $\ell=\tfrac12,\ldots,\nu q_a-\tfrac12$, and their
shifted momenta sum to $-(\nu q_a)^2/2$.  The corresponding
finite sum for $\nu q_a<0$ gives the same quadratic term.
We obtain
\bseq
\beq
\oR_\nu\of^a\oR_{-\nu}
&=\of^a+\nu s_aq_a,
\\[4pt]
\oR_\nu\oh^a\oR_{-\nu}
&=\oh^a-(q_a)^2\big(\nu\oA+\tfrac12\nu^2\big).
\label{eq:LRGTfh}
\eeq
\eseq
Since $\oA\mapsto\oA+\nu$,
we can define the invariant operators
\bseq
\label{eq:renfh}
\beq
(\orho^a_k)^\mathrm{inv}
&\equiv\orho^a_k-s_aq_a\oA\de_{k,0},
\label{eq:rhoinv}
\\[3pt]
(\of^a)^\mathrm{inv}
&\equiv\of^a-s_aq_a\oA,
\label{eq:renf}
\\[3pt]
(\oh^a)^\mathrm{inv}
&\equiv\oh^a+\tfrac12(q_a)^2\oA^2.
\label{eq:hinv}
\eeq
\eseq
This is the minimal LRGT invariant `dressing', with 
additive constants chosen to vanish in the Dirac vacuum at $A=0$.%
\footnote{
This still allows periodic functions of $\oA$, but 
these would give nonlocal operators with nontrivial dependence
on the Wilson loop parameterized by $\oA$.
If factors of $\hbar$ are restored, the terms arising from the
anticommutators are quantum corrections that vanish in the
classical limit.}
In our approach, the dressed operators are obtained without
requiring any regulator.
However, it is worth mentioning that the same result can be
obtained by gauge invariant
regularization and renormalization of the corresponding fermion
bilinears \cite{Manton:1985jm}.

We can now define the quantum versions of $H_\mathrm{gauge}$
and $H_\mathrm{ferm}$ in \Eqs{gaugekinetic} and \eq{Hferm}:
\bseq
\label{eq:HQ}
\beq
\oH_\mathrm{gauge}
&=\frac{g^2}{4\pi}(\oPi-\th)^2,
\label{eq:Hgauge}
\\[6pt]
\oH_\mathrm{ferm}
&=\sum_a(\oh^a)^\mathrm{inv}
=\sum_a\oh^a+\tfrac12q^2\oA^2,
\label{eq:HfermA1}
\eeq
\eseq
where $q^2\equiv\sum_a(q_a)^2>0$ and
\bseq
\beq
\sum_a\oh^a
&=\sum_{a,\ell}|\ell|\,\on^a_\ell
-\oA\,\op{\tilde Q},
\label{eq:HfermA2}
\\[3pt]
\on^a_\ell
&\equiv(\oa^a_\ell)^\dagger\oa^a_\ell,
\\[7pt]
\op{\tilde Q}
&\equiv\sum_a s_aq_a\of^a
=\sum_{a,\ell}\mathrm{sgn}(\ell)q_a\on^a_\ell.
\eeq
\eseq
Here $\on^a_\ell$ is the fermion occupation number operator
relative to the Dirac sea:
\beq
\on^a_\ell\ket{A,n}=n^a_\ell\ket{A,n}
\qquad\text{with}\qquad
n^a_\ell=0,1.
\eeq
The ordinary normal ordered axial charge transforms as
\beq
\oR_\nu\op{\tilde Q}\oR_{-\nu}
=\op{\tilde Q}+\nu q^2.
\label{eq:AxialLRGT}
\eeq
This demonstrates that the
quantum gauge and fermion Hamiltonians are separately
invariant under LRGTs.  It remains to define the Coulomb
interaction.

\subsection{The Coulomb Interaction}
\label{sec:Coulomb}
The classical Coulomb interaction \Eq{HCoulC} is a product of
local currents coupled by the nonlocal Green's function $K$.
Our quantization prescription therefore suggests 
that the quantum operator should be written as
a product of
the individually normal ordered current operators defined in
\sec{Quant}. We will see that this prescription yields
an operator whose
matrix elements are UV divergent, and therefore requires
a subtraction.
In position space, we therefore write
\beq
\oH_\mathrm{Coul}
&=\frac{g^2}{2}\int\!\D x\,\D y\,
\oJ^0(x)K(x-y)\oJ^0(y)
+\text{const}\cdot\id.
\label{eq:HCoulQx}
\eeq
Equivalently, in momentum space,
\beq
\oH_\mathrm{Coul}
&=\frac{g^2}{4\pi}\sum_{k\ne0}\frac{1}{k^2}
\oJ^0_{-k}\oJ^0_k
+\text{const}\cdot\id.
\eeq
The Green's function omits the zero mode, so the dressings in
\Eq{rhoinv} do not enter this expression.  Each current in the
product has finite matrix elements, but their product summed
over momenta has a UV divergence.  The subtraction constant is
the counterterm needed to remove this divergence:

\bseq
\label{eq:CoulombDivergence}
\beq
\sum_{k = 1}^{k_\mathrm{max}} \frac{\bra{0} \oJ^0_{k}
\oJ^0_{-k} \ket{0}}{k^2}
&= \sum_{k = 1}^{k_\mathrm{max}} \frac{q_+^2}{|k|}
\sim q_+^2 \log k_\mathrm{max},
\\[6pt]
\sum_{k = 1}^{k_\mathrm{max}} \frac{\bra{0} \oJ^0_{-k}
\oJ^0_{k} \ket{0}}{k^2}
&= \sum_{k = 1}^{k_\mathrm{max}} \frac{q_-^2}{|k|}
\sim q_-^2 \log k_\mathrm{max},
\eeq
\eseq
where 
\beq
q_\pm^2\equiv\sum_{a:s_a=\pm}(q_a)^2 > 
0
\qquad\text{and}\qquad
q_+^2+q_-^2=q^2.
\eeq
We choose the counterterm to be the vacuum subtraction:
\beq
\oH_\mathrm{Coul}
&\equiv \frac{g^2}{4\pi} \sum_{k\ne 0} 
\frac{\oJ^0_{-k} \oJ^0_k
- \bra{0} \oJ^0_{-k} \oJ^0_k \ket{0}}{k^2} ,
\label{eq:HCoulQ}
\eeq
where $\ket{0}$ denotes the fermion Fock vacuum.%
\footnote{The ordering in \Eq{HCoulQ} is sometimes called
current-algebra normal ordering.  It is different from fermion
normal ordering of the entire product,
$\no{\oJ^0_{-k}\oJ^0_k}$.}
More explicitly, the matrix elements of the Coulomb term 
in the truncated theory are given by
\beq
\bra{A', n'} \oH_\mathrm{Coul} \ket{A,n}
&= \frac{g^2}{4\pi} \sum_{k \neq 0} \frac{1}{k^2} 
\Big[ 
\bra{A', n'}\oJ^0_{-k} \oJ^0_k \ket{A,n}
- \de_{n'n} 
\bra{A',0} \oJ^0_{-k} \oJ^0_k \ket{A,0}
 \Big].
\eeq
The operator products in this expression are 
evaluated in the full Fock space,
with only the external states restricted to the truncated basis.

A nice property of this expression is that 
the sum over $k$ in the Coulomb term is a finite sum,
as we now demonstrate.
The truncation retains states with energies 
up to $E_\mathrm{max}$
under the free-fermion Hamiltonian
\beq
\oH_{0,\mathrm{ferm}}
\equiv\sum_{a,\ell}s_a\ell\,
\no{\opsi^{a\dagger}_\ell\opsi^a_\ell}
=\sum_{a,\ell}|\ell|\,\on^a_\ell.
\label{eq:H0ferm}
\eeq
To see why the momentum sum in the Coulomb term becomes finite,
introduce the normalized current oscillators for each chirality
$s=\pm$ with $q_s^2>0$:
\beq
\oB_{s,j}
\equiv\frac{1}{\sqrt{j q_s^2}}
\sum_{a:s_a=s}q_a\orho^a_{sj}
\qquad\text{with}\qquad
j=1,2,\ldots\,.
\label{eq:alphabetadefn}
\eeq
(Every anomaly-free gauge theory has $q_\pm^2 > 0$.)
These are precisely the oscillators in the gauge-charge directions
that appear in the bosonized description in \sec{Bosonization}.
The current algebra, given explicitly in \Eq{rhocomm}, implies
\beq
[\oB^{\vphantom \dagger}_{s,j},\oB^\dagger_{s',j'}]
=\de_{ss'}\de_{jj'}
\qquad\text{and}\qquad
[\oB^{\vphantom \dagger}_{s,j},\oB_{s',j'}]=0.
\eeq
They preserve every fermion number and obey
\beq
[\oH_{0,\mathrm{ferm}},\oB_{s,j}]
=-j\oB_{s,j}
\qquad\text{and}\qquad
[\oH^{\vphantom \dagger}_{0,\mathrm{ferm}},\oB^\dagger_{s,j}]
=j\oB^\dagger_{s,j}.
\eeq
Thus they are bosonic annihilation and creation operators acting
directly on the fermion Fock space.
Because $\oH_{0,\mathrm{ferm}}$ has a nonnegative spectrum,
$\oB_{s,j}$ annihilates every state in the truncated basis for
$j>E_\mathrm{max}$.
Combining the positive and negative momentum terms in \Eq{HCoulQ},
the vacuum subtraction gives
\beq
\oH_\mathrm{Coul}
&=\frac{g^2}{2\pi}\sum_{j=1}^\infty\frac1j
\bigg[
\underbrace{
q_+^2\oB^\dagger_{+,j}\oB^{\vphantom \dagger}_{+,j}
+q_-^2\oB^\dagger_{-,j}\oB^{\vphantom \dagger}_{-,j}
}_{\De H_{0,\mathrm{ferm}}=0}
\nn
&\qquad\qquad
+\sqrt{q_+^2q_-^2}\Big(
\underbrace{\oB^{\vphantom \dagger}_{+,j}\oB^{\vphantom \dagger}_{-,j}}_
{\De H_{0,\mathrm{ferm}}=-2j}
+\underbrace{\oB^\dagger_{+,j}\oB^\dagger_{-,j}}_
{\De H_{0,\mathrm{ferm}}=2j}
\Big)\bigg].
\label{eq:HCoulOscillators}
\eeq
For matrix elements between states in the truncated basis,
the terms in the first line vanish for $j>E_\mathrm{max}$
because $\oB_{\pm,j}$ annihilates the state.
The terms in the second line change the free-fermion energy
by $\pm2j$, and therefore vanish for
$j>\frac12 E_\mathrm{max}$.

In \sec{Bosonization}, we will use the bosonized description
to show that the prescription
\Eq{HCoulQ}, together with the covariantized fermion Hamiltonian,
gives a Lorentz invariant theory when the gauge anomaly cancels.
The vacuum subtraction fixes the additive energy constant by
requiring $\bra0\oH_\mathrm{Coul}\ket0=0$.
In the following section, we will also show that this Hamiltonian
correctly reproduces the chiral anomalies.

We have now constructed the full quantum Hamiltonian:
\beq
\label{eq:Htotal}
\oH=\oH_\mathrm{gauge}+\oH_\mathrm{ferm}
+\oH_\mathrm{Coul}.
\eeq
Its three terms are each invariant under LRGTs,
and matrix elements of these terms are finite.

\section{Chiral Anomalies}
\label{sec:anomaly}
We now show that the gauge-fixed quantum theory defined above reproduces
the expected chiral anomalies.
The anomaly is obtained without introducing a UV regulator.
We will see that the requirement of invariance under LRGTs fixes the dressing of the
normal ordered charges, and the dressed charges need not be conserved.
In particular, we show that we cannot impose $Q = 0$
in the quantum theory unless the gauge anomaly vanishes.

\subsection{Charge Conservation}

We define a general normal ordered $U(1)$ charge by (see \Eq{fhat})
\beq
\oQ_r \equiv r \cdot \of = \sum_a r_a \of^a
\qquad\text{with}\qquad
r_a \in \mathbb{Z},
\label{eq:oQrdef}
\eeq
so the normal ordered gauge charge is given by $\oQ = \oQ_q$ in this notation.
We restrict ourselves to integer linear combinations of fermion numbers,
so that these charges have integer eigenvalues.
We will see that the dressing required by LRGT invariance can spoil
this property.

Applying \Eq{renf}, we have
\beq
\label{eq:Qranomaly}
\oQ_r^\mathrm{inv} \equiv \sum_a r_a (\of^a)^\mathrm{inv}
= \oQ_r - \scr{A}_r \oA,
\eeq
where
\beq
{\cal A}_r \equiv \sum_a s_a r_a q_a
= r \cdot \tilde{q}
\qquad\text{with}\qquad
\tilde{q}_a \equiv s_a q_a,
\label{eq:defcalA}
\eeq
is the coefficient for the mixed anomaly between the charge $Q_r$ and the gauge
charge $Q$.
(For $r = q$, it is the gauge anomaly.)
Indeed, under an LRGT we have
\beq
\oR_\nu \oQ_r \oR_{-\nu} = \oQ_r + \nu \scr{A}_r,
\eeq
so the $\oA$ term in \Eq{Qranomaly} is precisely the dressing that
makes the charge invariant.
The normal ordered charge simply adds up the 
charge-weighted fermion numbers of the excited states relative to the Dirac vacuum: 
\beq
\oQ_r \ket{A, n} = Q_r(n) \ket{A,n}
\qquad\text{with}\qquad
Q_r(n) \equiv \sum_{a,\ell} s_a\,\mathrm{sgn}(\ell)
r_a n^a_\ell \in \mathbb{Z}.
\label{eq:QrDef}
\eeq
This is conserved by the time evolution because
\beq
\big[ \oQ_r, \oH \big] = 0,
\eeq
for any choice of $r$.
On the other hand, if $\scr{A}_r \ne 0$,
$\oQ_r$ is not invariant under LRGTs,
and the dressed gauge invariant charge $\oQ_r^\mathrm{inv}$ is not conserved
(and is not integer-valued):
\beq
\d_0 \oQ_r^\mathrm{inv} = -\scr{A}_r \d_0 \oA
= -\frac{\scr{A}_r}{2\pi} \int \! \D x\ggap \oF_{01}.
\label{eq:IntGaugeAnom}
\eeq
This is the standard integrated form of the chiral anomaly.

This reveals the inconsistency of a theory that has a nonvanishing gauge anomaly. The constraint $Q = 0$ in the classical theory implies that the quantum theory must have
\beq
\label{eq:Gauss}
\oQ^\mathrm{inv} \ket{\Psi}_\mathrm{phys} = 0,
\eeq
where $\ket{\Psi}_\mathrm{phys}$ are physical gauge invariant states (see \sec{GaugeInvStates} below).
However, if $\scr{A}_q \ne 0$, then $\oQ^\mathrm{inv}$ is not conserved, and this condition is inconsistent with time evolution. It is an important check that the obstruction arising from the gauge anomaly is clear in the gauge-fixed theory before truncation. Furthermore, this condition will still be valid in the truncated theory.

It is also interesting to consider the normal ordered momentum operator:
\beq
\oP &= \sum_a s_a\oh^a
= \sum_{a,\ell}\ell\,\no{\opsi^{a\dagger}_\ell\opsi^a_\ell}
- \oA\,\oQ.
\eeq
Using \Eq{hinv}, its
gauge invariant dressing is
\beq
\oP^\mathrm{inv} \equiv \sum_a s_a(\oh^a)^\mathrm{inv}
= \oP + \tfrac12\scr{A}_q\oA^2.
\eeq
Provided the gauge anomaly vanishes, $\oQ^\mathrm{inv}=\oQ$ and
$\oP^\mathrm{inv}=\oP$.
For an anomaly-free theory, $\oP$ is conserved on the
physical subspace:
\beq
\d_0\oP \ket{\Psi}_\mathrm{phys} = 0
\eeq
because $\oQ \ket{\Psi}_\mathrm{phys} = 0$.
Again, note that no UV regulator is required to obtain these
results.

\subsection{Current Conservation}
We can also obtain the local form of the chiral anomaly
using this language.
For this we need the commutation relations
\cite{Sugawara:1967rw}
\beq
\big[ \orho^a_k,\, \orho^b_{k'} \big]
= s_a k \de^{ab} \de_{k+k'\!,0}.
\label{eq:rhocomm}
\eeq
This can be verified by normal ordering,
which involves only a finite number of nonzero
anticommutators.
Therefore, no regulator is required to obtain the result.

We define the dressed current (compare \Eq{J0kcl}) by
\beq
\oJ^{0,\mathrm{inv}}_{r,k}
= \begin{cases}
\displaystyle
\sum_a r_a \orho^a_k & k \ne 0,
\\[16pt]
\oQ^\mathrm{inv}_r & k = 0.
\end{cases}
\eeq
The nonzero modes are already invariant under LRGTs, so only the zero
mode needs dressing.
We restrict attention to $k \ne 0$, since the anomaly
equation for $k = 0$ is the integrated anomaly \Eq{IntGaugeAnom} discussed
above.
Noting that $[\oH_\text{gauge}, \oJ^{0,\mathrm{inv}}_{r,k}] = 0$, we then compute
\beq
\d_0 \oJ^{0,\mathrm{inv}}_{r,k} &= \ii \big[\oH_\mathrm{ferm},\,
\oJ^{0,\mathrm{inv}}_{r,k} \big]
+ \ii \big[\oH_\mathrm{Coul},\,
\oJ^{0,\mathrm{inv}}_{r,k} \big].
\eeq
We obtain
\beq
\big[\oH_\mathrm{ferm},\,
\oJ^{0,\mathrm{inv}}_{r,k} \big]
= -k \oJ^{1,\mathrm{inv}}_{r,k},
\eeq
where
\beq
\oJ^{1,\mathrm{inv}}_{r,k}
= \sum_a s_a r_a \orho^a_k
\qquad
(k \ne 0).
\eeq
Using \Eq{rhocomm}, we obtain
\beq
\big[\oH_\mathrm{Coul},\,
\oJ^{0,\mathrm{inv}}_{r,k} \big]
&=-\frac{g^2}{2\pi k}\, {\cal A}_r
\op{J}^{0,\mathrm{inv}}_{q,k}
\qquad (k \ne 0).
\label{eq:JHCcommcheck}
\eeq
For $k \ne 0$ we identify
\beq
\frac{\ii g^2}{k}
\oJ^{0,\mathrm{inv}}_{q,k} = \oF_{01,k} = \int \D x\, e^{-\ii k x}\,
\oF_{01}(x).
\eeq
This follows from the fact that the gauge invariant electric
field $\oF_{01,k}$ must satisfy the $A_0$ equation of motion
$\d_1\oF_{01}=-g^2\oJ^{0,\mathrm{inv}}_q$.
The zero mode of the spatial current is similarly dressed as
$\oJ^{1,\mathrm{inv}}_{r,0}=\sum_a s_a r_a(\of^a)^\mathrm{inv}$.

We therefore obtain the standard anomaly equation
in position space:
\beq
\d_\mu \oJ^{\mu,\mathrm{inv}}_r(x) = -\frac{\scr{A}_r}{2\pi} \oF_{01}(x),
\eeq
where the zero mode of this equation comes from \Eq{IntGaugeAnom}.
This shows
that the anomaly is completely fixed by
LRGT invariance and the current algebra \Eq{rhocomm};
in particular, it is obtained without any explicit UV regularization.

\section{Gauge Invariant States}
\label{sec:GaugeInvStates}

In this section, we define the gauge invariant states and give a formula for matrix elements of gauge invariant operators in a form suitable for Hamiltonian truncation.  
The construction below is equivalent to classic Hamiltonian treatments of the Schwinger model on a circle \cite{Manton:1985jm,Hetrick:1988yg,Iso:1988zi} and its generalization to anomaly-free Abelian chiral gauge theories \cite{GrosseLangmann2000}.
Our discussion attempts to clarify the conceptual issues (in particular the physical theta parameter), and also give a simple concrete expression for the matrix elements of the Hamiltonian in physical states that is suitable for the numerical work below.

We assume that the theory is anomaly-free, so that we can consistently impose Gauss' law.
We work in the Hilbert space of states satisfying Gauss' law
\beq
\scr{H} = \big\{ \ket\Psi : \op{Q} \ket\Psi = 0 \big\}.
\eeq
The space of states is a tensor product
$\scr{H} = \scr{A} \otimes \scr{F}$,
where $\scr{A}$ consists of the wavefunctions of $A$, and $\scr{F}$ is the fermion Fock space
\beq
\scr{F} = \text{span} \big\{ \ket{n} : Q(n) = 0 \big\}.
\eeq
The LRGT split the Fock space into orthogonal subspaces:
\beq
\scr F = \bigoplus_{\mu\in\ZZ} \scr{F}_\mu
\qquad \text{with}\qquad
\oR_\nu \scr{F}_\mu = \scr{F}_{\mu - \nu} \perp \scr{F}_\mu
\quad\text{for}\quad \nu \neq 0.
\eeq
The spaces $\scr{F}_\mu$ are labeled using the normal-ordered axial charge
\beq
\op{\tilde Q}\ket{A,n}
=\tilde Q(n)\ket{A,n}
\qquad\text{where}\qquad
\tilde Q(n)=\tilde q\cdot f(n)
=\sum_a s_a q_a f^a(n),
\eeq
so $\scr{F}_0$ contains the Fock vacuum $\ket{0}$.
The LRGT $\hat R_\nu$ shifts the axial charge by $-\nu q^2$, so we have
\beq
{\scr F}_\mu
= \text{span}\big\{
\ket{n} : Q(n) = 0,\ \mu q^2 \le \tilde{Q}(n) \le (\mu+1)q^2 - 1 \big\}.
\eeq

We can choose a normalizable basis for $\scr{A}$ (for example simple harmonic oscillator wavefunctions) so that $\scr{H}$ has a basis of normalizable states.
Then $\scr H$ has a corresponding orthogonal decomposition
\beq
\scr{H} = \bigoplus_{\mu \in \ZZ} \scr{H}_\mu
\qquad \text{with}\qquad
\scr{H}_\mu = \scr{A} \otimes \scr{F}_\mu
\eeq
with $\oR_\nu \scr{H}_\mu = \scr{H}_{\mu - \nu}$, {\it etc\/}.
Any state $\ket\Psi \in \scr{H}$ can be uniquely decomposed as
\beq
\ket\Psi = \sum_\mu \ket{\Psi_\mu}
\qquad \text{with}\qquad
\ket{\Psi_\mu} \in \scr{H}_\mu.
\eql{Psidecomp}
\eeq
In the construction below, we restrict attention to
states for which the sum over $\mu$ in this decomposition is finite.
These states are dense in $\scr H$.

The states within a given orthogonal subspace $\scr{H}_\mu$ can be labeled by the eigenvalues of the gauge-invariant operator
\beq
\eql{Xdef}
\op{X}^{\rm inv} \equiv e^{2\pi i \op{\tilde{Q}} / q^2}.
\eeq
Note that this is written in terms of the normal-ordered charge $\op{\tilde{Q}}$, but it is gauge invariant because 
$\tilde{Q} \mapsto \tilde{Q} - \nu q^2$ under an LRGT.
This will be useful in labeling the physical states.

We want the physical states of our theory to be gauge invariant.
However, the only gauge invariant state in $\scr{H}$ is the zero state.
To see this consider an invariant state $\ket\Psi$ and decompose it as in \Eq{Psidecomp}.
We have $\ket{\Psi_\mu} = \oR_{-\mu} \ket{\Psi_0}$ for some $\ket{\Psi_0} \in \scr{H}_0$, and we have
\beq
\big\lVert \ket{\Psi} \big\rVert^2
= \sum_{\nu \in \ZZ} \big\lVert \ket{\Psi_0} \big\rVert^2,
\eeq
which diverges unless $\ket{\Psi_0} = 0$.
This is a consequence of the `Hilbert Hotel' nature of the LRGT.
The LRGT relate infinitely many mutually orthogonal sectors, and
invariance under LRGT requires the components in all these
sectors to have equal norm.
The norm of the full state is therefore finite only if every
orthogonal component vanishes.

This means that we cannot identify the physical states with invariant states in $\scr H$.
Instead, we define the LRGT invariant states as equivalence classes of states in $\scr H$ related by LRGT.
If $\ket\Psi \sim \oR_\nu \ket\Psi$, then we must have $\ket\Psi - \oR_\nu\ket\Psi \sim 0$.
We therefore define the equivalence relation by
\beq
\ket\Psi \sim \ket{\Psi'}
\qquad\Leftrightarrow\qquad
\ket\Psi - \ket{\Psi'} \in \scr N,
\eeq
where
\beq
\scr N = \text{span}\big\{ (\oR_\nu - 1)\ket\Psi
: \nu\in\ZZ,\ \ket\Psi \in \scr H \big\}
\eeq
is the subspace of states in $\scr H$ that are equivalent to the zero state.%
\footnote{We could introduce an arbitrary phase into the definition of $\scr N$ by defining the null state by
\beq
\scr{N}_\al = \text{span}\big\{ (e^{i \nu \al} \oR_\nu - 1)\ket\Psi
: \nu\in\ZZ,\ \ket\Psi \in \scr H \big\},
\eeq
but this is equivalent to redefining $R_\nu \to e^{i\nu\al} R_\nu$.
We are already allowing an arbitrary phase in $R_\nu$ (see \Eq{Rnudefn}), so this does not introduce any additional independent parameter in the construction.
When we define the action of gauge invariant operators on physical states, we will see that this involves an arbitrary phase that contributes to $\bar\th$.}
The space $\scr N$ is a vector space called the `null space' \cite{Gupta:1949rh, Bleuler:1950cy}.
The physical states are identified with the equivalence classes defined by this relation.
In other words, the physical states are elements of the quotient vector space
\beq
\scr{H}_\mathrm{phys} = \scr{H}/\scr{N}.
\eeq
We will denote the physical states by $\ket{\Psi}_\mathrm{phys}$, where the label $\Psi$ denotes a state $\ket\Psi \in \scr H$.

Next, we choose a representative of each equivalence class by choosing a representative in $\scr{H}_0$.
Given a state $\ket\Psi \in \scr{H}$, we use the decomposition \Eq{Psidecomp} to define
\beq
\op{T} \ket\Psi \equiv \sum_\mu
\oR_\mu \ket{\Psi_\mu} \in \scr{H}_0.
\eeq
$\op T$ is a linear map, so it preserves the vector space structure on the physical states.
The operator $\op T$ allows us to identify the space of physical states with $\scr{H}_0$: we are `fixing the gauge' by choosing a representative of every physical state in $\scr{H}_0$.

We then define the inner product on physical states by 
\beq
{}_\mathrm{phys}\braket{\Phi}{\Psi}_\mathrm{phys}
\equiv \left( \op{T} \ket\Phi \right)^\dagger \op T \ket{\Psi},
\eeq
where the inner product on the \rhs\ is the inner product in $\scr H$.
To show that this inner product is positive-definite, we must show that
$\op T \ket\Psi = 0$ if and only if $\ket\Psi \in \scr{N}$, or
\beq
\ker \op T = \scr N.
\eeq
To prove this, first note that
\beq
\op T \oR_\nu = \op T,
\eeq
so $\ket\Psi \in \scr N$ implies $\op T\ket\Psi = 0$.
Conversely, 
\beq
\ket\Psi-\op T\ket\Psi
=\sum_\mu(\oR_{-\mu}-1)\oR_\mu\ket{\Psi_\mu}
\in\scr N,
\eeq
so $\op T\ket\Psi = 0$ implies $\ket\Psi \in \scr N$.
This shows that ${}_\mathrm{phys}\braket{\Psi}{\Psi}_\mathrm{phys} = 0$ if and only if $\ket\Psi \in \scr N$ (or $\ket\Psi \sim 0$).
The other standard properties of an inner product are inherited from the inner product on $\scr{H}$.

Next, we consider the action of operators on physical states.
The operators must be gauge invariant:
\beq
\oR_\nu \osO \oR_{-\nu} = \osO.
\eeq
A general gauge invariant operator will mix the different orthogonal subspaces.
The operators of interest are those that map $\scr{H}_\mu$ to $\scr{H}_{\mu-\nu}$ for some definite $\nu \in \mathbb{Z}$:
\beq
\osO_\nu \scr{H}_\mu \subset \scr{H}_{\mu-\nu}.
\eeq
(The LRGT gauge transformation $\oR_\nu$ is an example of such an operator.)
Gauge invariant operators map null states to null states:
\beq
\hat{\scr O}_\nu(R_\la-1) \ket\Psi
=(R_\la-1)\hat{\scr O}_\nu \ket\Psi
\in\scr N.
\eeq
Therefore gauge-equivalent input states give gauge-equivalent outputs.
We define the action of these operators by acting with $\osO_\nu$ on a representative in $\scr{H}_0$ and then returning the result to $\scr{H}_0$:
\beq
\osO_\nu \ket{\Psi_0}_\mathrm{phys}
\equiv e^{-i\nu\th_\mathrm{J}} \ket{\Xi_0}_\mathrm{phys}
\qquad\text{and}\qquad
\ket{\Xi_0} = \oR_{-\nu} \osO_\nu \ket{\Psi_0} \in \scr{H}_0.
\eeq
Here $\th_\mathrm{J}$ is an arbitrary phase that we call the `Jackiw phase'~\cite{Jackiw:1983nv}.
Equivalently, the matrix elements of the operator $\osO_\nu$ are given by
\beq
{}_\mathrm{phys}\bra{\Phi_0} \osO_\nu
\ket{\Psi_0}_\mathrm{phys}
= e^{-i\nu\th_\mathrm{J}} \bra{\Phi_0} \oR_{-\nu} \osO_\nu \ket{\Psi_0}
\qquad \text{with}\qquad
\ket{\Psi_0}, \ket{\Phi_0} \in \scr{H}_0.
\eql{PhysMatrixElements}
\eeq
Note that with this definition,
\beq
\oR_\nu \ket{\Psi_0}_\mathrm{phys} = e^{-i\nu\th_\mathrm{J}}
\ket{\Psi_0}_\mathrm{phys}.
\eeq
That is, the physical states are invariant under the action of a LRGT up to a phase.

With these definitions, we can now compute the matrix elements of the Hamiltonian.
The Hamiltonian is invariant under LRGT.
We first consider the case where the Hamiltonian contains only gauge interactions.
In this case, we have
\beq
\big[\oH, \op{\tilde{Q}} \big] = 0,
\eeq
where $\op{\tilde{Q}}$ is the normal-ordered axial charge operator,
so $\oH$ maps $\scr{H}_0$ to itself.
Therefore, no LRGT is required to compute the matrix elements of the Hamiltonian,
and we have
\beq
{}_\mathrm{phys}\bra{\Phi_0} \oH \ket{\Psi_0}_\mathrm{phys}
= \bra{\Phi_0} \oH \ket{\Psi_0}
\qquad \text{with}\qquad
\ket{\Phi_0}, \ket{\Psi_0} \in \scr{H}_0.
\eeq
That is, we simply evaluate the matrix elements of the Hamiltonian in $\scr{H}_0$.
Explicitly, a complete set of matrix elements of the Hamiltonian is given by
\beq
\bra{A', n'} \oH \ket{A, n}
\qquad\text{with}\qquad
0 \le \tilde{Q}(n), \tilde{Q}(n') \le q^2 - 1.
\eeq

The Hamiltonian depends on the angle $\th$ through the gauge kinetic term \Eq{gaugekinetic}.
However, this dependence can be removed by redefining the physical states by 
\beq
\ket{\Psi}_\mathrm{phys} \to \oV_\th \ket{\Psi}_\mathrm{phys}
\qquad\text{with}\qquad
\oV_\th \equiv e^{-i\th \op{\tilde{Q}}^\mathrm{inv}/q^2},
\label{eq:RedefStateV}
\eeq
where $\op{\tilde{Q}}^\mathrm{inv}$ is the gauge invariant (`dressed') axial charge operator defined in
\Eq{Qranomaly} with $r = \tilde{q}$.
We have
\beq
\oV_\th = e^{i\th\hat A}
 e^{-i\th {}:{}\hat{\tilde Q}{}:{}/q^2},
\eeq
and hence the gauge kinetic term becomes
\beq
\oV_\th^\dagger \big(\oPi - \th\big)^2 \oV_\th = 
e^{-i\th\hat A} \big(\oPi - \th\big)^2  e^{i\th\hat A}
= \oPi^2.
\eeq
The remaining terms in the Hamiltonian are unaffected, so the matrix elements of $\oH$ are independent of $\th$ in the redefined basis.
This makes it manifest that the energy spectrum is independent of $\th$.
This is as expected, because the Hamiltonian describes massless fermions and has no explicit chiral symmetry breaking.

The results above are sufficient to obtain the numerical results for the massless Schwinger model and the 345 model presented in \sec{Numerics}.
However, motivated by recent claims that 4D gauge theories contain states with different theta angles \cite{Kaplan:2025bgy}, we discuss more general theories with explicit chiral symmetry breaking and nontrivial $\th$ dependence.

\subsection{Fermion Masses and $\bar{\theta}$}
\label{sec:theta}
We will give an example with nontrivial $\th$ dependence using the massive one-flavor Schwinger model, which has 2 Weyl fermions with  $q_+ = q_- = 1$.
The Hamiltonian contains the mass term
\beq
\De\oH_\mathrm{mass} = m\gap \osO_\mathrm{mass}
+ \hc,
\eeq
where 
\beq
\osO_\mathrm{mass} = \int\! \text{d}x\, \opsi_+^\dagger\opsi_-.
\eeq
The mass term is invariant under LRGT and commutes with $\oA$,
but it has nonzero normal-ordered axial charge:
\beq
\big[\op{\tilde Q},\osO_\mathrm{mass}\big]
= 2\osO_\mathrm{mass}.
\eql{massAxialCharge}
\eeq
Gauss' law implies 
$Q(n) = f_+(n)+f_-(n)=0$,
so $\tilde Q(n)=f_+(n)-f_-(n)=2f_+(n)$ is even.
In this theory $q^2 = 2$, so the states in $\scr{H}_0$ have $\tilde{Q}(n) = 0$.

The physical matrix elements of the mass term are therefore given by
(see \Eq{PhysMatrixElements})
\beq
{}_\mathrm{phys}\bra{\Phi_0} 
\osO_\mathrm{mass}
\ket{\Psi_0}_\mathrm{phys}
= e^{\ii\th_\mathrm{J}} \bra{\Phi_0} 
\oR_1\osO_\mathrm{mass}
\ket{\Psi_0}.
\eeq
Making the state redefinition \Eq{RedefStateV}, we have
\beq
\op{V}_\th^\dagger\osO_\mathrm{mass}\op{V}_\th
&= e^{\ii\th{}:{}\op{\tilde Q}{}:{}/2}
\osO_\mathrm{mass}
e^{-\ii\th:{}\op{\tilde Q}{}:{}/2}
= e^{\ii\th}\osO_\mathrm{mass},
\eql{massVTheta}
\eeq
and hence
\beq
{}_\mathrm{phys}\bra{\Phi'_0} 
m \osO_\mathrm{mass}
\ket{\Psi'_0}_\mathrm{phys}
&= m e^{\ii(\th + \th_\mathrm{J})} 
\bra{\Phi_0} \oR_1 \osO_\mathrm{mass} \ket{\Psi_0},
\eeq
where $\ket{\Psi'_0} = \op{V}_\th \ket{\Psi_0}$, {\it etc\/}.
We have used the fact that $\oR_1$ commutes with $\oV_\th$, which holds because $\oV_\th$ is gauge invariant.
In the redefined basis, the remaining terms in the Hamiltonian are independent of $\th$, as discussed above.
For $m\neq 0$, the angles and the phase of the mass therefore enter
the Hamiltonian through the combination
\beq
\bar\th = \th+\th_\mathrm{J}+\arg m
\pmod{2\pi}.
\eql{barTheta}
\eeq
This is the physical theta angle.

In this construction, it is clear that $\bar\th$ is a parameter of the theory, rather than a property of physical states that depends on the state:
$\th$ and $\arg(m)$ are couplings in the Hamiltonian, 
and $\th_\mathrm{J}$ is a parameter in the definition of the action of gauge invariant operators on physical states.
\Ref{Kaplan:2025bgy} argues that the theory must contain states with different theta angles in order to be consistent with the superposition principle.
The construction above gives an explicit counterexample to this argument: arbitrary linear combinations of physical states are allowed, and all physical states have the same value of $\bar\th$.
\Ref{Kaplan:2025bgy} instead write the Hamiltonian acting on wavefunctions without identifying the large residual gauge transformations, decompose its states into sectors labeled by their phase under large gauge transformations ($\th_\mathrm{J}$ in our construction), and treat arbitrary linear combinations of these sectors as physical states.
However, such superpositions are not gauge invariant, even up to an overall phase, and therefore are not physical states.
Indeed, the requirement
\beq
\big( \oR_\nu \ket{\Psi} \big)_\mathrm{phys}
= e^{-i\nu\th_\Psi} \ket{\Psi}_\mathrm{phys}
\eeq
is compatible with taking superpositions of physical states only if $\th_\Psi$ is the same (mod $2\pi$) for all physical states, as in our construction.
This requirement follows from the assumption that residual gauge transformations represent redundancies of the theory in finite volume.

\section{Bosonization}
\label{sec:Bosonization}
In this section, we construct the exact solution of massless
anomaly-free two-dimensional Abelian chiral gauge theories using 
bosonization~\cite{Mattis:1964wp,Coleman:1974bu,Mandelstam:1975hb}.
Our construction uses the $U(1)$ current algebra and the corresponding 
Sugawara expression for the free-fermion Hamiltonian 
\cite{Sugawara:1967rw}.
Exact results for general chiral gauge theories were obtained
in Refs.~\cite{Banks:1981sc,Halliday:1985tg};
the complete physical spectrum was explicitly obtained 
in Ref.~\cite{KutasovSchwimmer1995}.
Here we give 
a Hamiltonian derivation on $S^1$
for an arbitrary $U(1)$ chiral gauge theory.
We show that the exact spectrum consists of one massive scalar,
$N_+-1$ massless right-moving chiral bosons, and
$N_--1$ massless left-moving chiral bosons.

\subsection{Free-Fermion Bosonization}
We start by describing the bosonization of a system of free fermions.
We define the normalized operators
\beq
\ob^a_j
\equiv \frac{1}{\sqrt j}\orho^a_{s_a j}
\qquad\text{and}\qquad
(\ob^a_j)^\dagger
\equiv \frac{1}{\sqrt j}\orho^a_{-s_a j}
\qquad\text{with}\qquad
j = 1, 2, 3, \ldots .
\label{eq:bososcillators}
\eeq
\Eq{rhocomm} implies that these have
the commutation relations of bosonic creation and
annihilation operators:
\beq
\begin{split}
&\big[\ob^a_j,(\ob^b_{j'})^\dagger\big]
=\de^{ab}\de_{jj'},\\[5pt]
&\big[\ob^a_j,\ob^b_{j'}\big]
=\big[(\ob^a_j)^\dagger,
(\ob^b_{j'})^\dagger\big]=0.
\end{split}
\label{eq:bososcillatoralgebra}
\eeq
They commute with every $\of^a$ and therefore preserve all
fermion numbers.
They act as raising and lowering operators for the free-fermion
energies:
\beq
\big[ \oh^a, \ob^b_j \big]
= -j \de^{ab} \ob^b_j
\qquad\text{and}\qquad
\big[ \oh^a, (\ob^b_j)^\dagger \big]
= j \de^{ab} (\ob^b_j)^\dagger.
\eeq
For each collection $f$ of integers $f^a$, there is a unique
state $\ket f$ with lowest free-fermion energy satisfying
\beq
\of^a\ket f=f^a\ket f,
\eeq
with
\beq
\ob^a_j\ket f=0
\quad\text{for all } a, j.
\label{eq:boschargedvacuum}
\eeq
The bosonic creation operators generate every fermion 
state in this charge sector.
Therefore, the fermionic Fock space has the structure of 
a direct sum of bosonic Fock spaces, one for each 
$f \in \mathbb{Z}^N$.

The free-fermion Hamiltonian can be written in terms
of these operators:
\beq
\oH_{0,\mathrm{ferm}} \equiv \sum_{a,\ell}s_a\ell\,
\no{\opsi^{a\dagger}_\ell\opsi^{a\vphantom{\dagger}}_\ell}
=\frac12\sum_a\big(\of^a\big)^2
+\sum_a\sum_{j=1}^\infty
j(\ob^a_j)^\dagger\ob^a_j
.
\label{eq:bosfreeHamiltonian}
\eeq
The first term on the right
is the energy of the state $\ket{f}$ and the remaining
terms are the free Hamiltonian of a set of independent bosonic oscillators for each momentum level $j$. 
This is the bosonized description of the free-fermion 
Hamiltonian. Note that the $\op{b}^a_j$ are invariant under LRGTs, while the $\op{f}^a$ is not.

\subsection{Gauge Zero Mode}
We now include the coupling between the fermions and the gauge zero mode. Combining \Eq{bosfreeHamiltonian} with the covariantized fermion
Hamiltonian \Eq{HQ} gives
\beq
\oH_\mathrm{ferm}
&=\oH_{0,\mathrm{ferm}}-\oA\sum_a s_a q_a\of^a
+\tfrac12q^2\oA^2
\nn
&=\frac12\sum_a
\big[(\of^a)^\mathrm{inv}\big]^2
+\sum_a\sum_{j=1}^\infty
j(\ob^a_j)^\dagger\ob^a_j,
\label{eq:bosfermionHamiltonian}
\eeq
where  we used \Eq{renf} in the last line.  
This is now manifestly invariant under LRGTs.

On states with fermion numbers $f$,
the zero-mode potential is
\beq
V_f(A)
&=\frac12\sum_a(f^a-s_a q_a A)^2
\nn
&=\frac12q^2(A-A_f)^2
+\frac12\left[
\sum_a(f^a)^2
-\frac{(\tilde q\cdot f)^2}{q^2}
\right],
\label{eq:boszeropotential}
\eeq
where
\beq
A_f\equiv\frac{\tilde q\cdot f}{q^2}
\qquad\text{and}\qquad
\tilde q_a=s_a q_a.
\eeq
Under an LRGT the state labels obey
$A\mapsto A-\nu$ and $f\mapsto f-\nu\tilde q$, so $A-A_f$ is
unchanged. Together with $\oH_\mathrm{gauge}$, the first term in
\Eq{boszeropotential} is a harmonic oscillator with frequency $\omega_0$, where
\beq
\omega_0^2=\frac{g^2q^2}{2\pi}.
\label{eq:bosmasszero}
\eeq
As shown in \sec{GaugeInvStates}, the gauge invariant states 
in the theory with only gauge interactions are obtained
by choosing a representative charge sector.
This means that each gauge invariant basis state is represented by a state
with a definite value of the fermion numbers $f$.

\subsection{The Coulomb Term}
To express the Coulomb interaction in terms of the bosonic
creation and annihilation operators, we define the annihilation
operators along the charge directions
\beq
\oB_{\pm,j} \equiv \frac{1}{\sqrt{q_\pm^2}} \sum_{a:s_a=\pm} 
q_a \ob^a_j,
\eeq
satisfying
\beq
\big[ \oB\sub{s,j},\, \oB^\dagger_{s',j'} \big]
= \de_{ss'}\de_{jj'}.
\eeq
In terms of this, we have
\beq
\oJ^0_j
=\sqrt{j q^2/2} \, \big( \oB_{+,j}
+ \oB^\dagger_{-,j}  \big)
\qquad\text{and}\qquad
\oJ^0_{-j}
=\sqrt{j q^2/2} \, \big(
\oB^\dagger_{+,j} + \oB\sub{-,j} \big).
\label{eq:boscurrentmodes}
\eeq
Only nonzero momentum modes appear in the Coulomb term,
so the normal ordered operators $\oJ^0_{\pm j}$ are invariant
under LRGTs without dressing.
The vacuum subtractions that appear in \Eq{HCoulQ} are given by
\beq
\bra{0}\oJ^0_{-j}\oJ^0_j\ket{0}=jq_-^2
\qquad\text{and}\qquad
\bra{0}\oJ^0_j\oJ^0_{-j}\ket{0}=jq_+^2.
\label{eq:boscurrentcontractions}
\eeq
Substituting \Eq{boscurrentmodes} and
\Eq{boscurrentcontractions} into the current-algebra ordering
prescription \Eq{HCoulQ}, and combining the terms with 
$k = \pm j$ gives
\beq
\oH_\mathrm{Coul}
&=\frac{g^2 q^2}{4\pi}\sum_{j=1}^\infty \frac1j
\bigg[
\oB^\dagger_{+,j}\oB\sub{+,j}
+ \oB^\dagger_{-,j} \oB\sub{-,j}
+ \oB_{+,j} \oB_{-,j}
+ \oB^\dagger_{+,j} \oB^\dagger_{-,j}
\bigg].
\label{eq:bosCoulombHamiltonian}
\eeq
The vacuum subtraction and operator ordering prescription
in \Eq{HCoulQ} are essential for this result.
In particular, the vacuum subtraction converts
the two diagonal products into the bosonically normal ordered number
operators in \Eq{bosCoulombHamiltonian}.
Thus the vacuum subtraction introduced in \sec{Coulomb}
is precisely the bosonic normal ordering of the current products,
even though each current was already normal ordered as a
fermion bilinear.

\Eqs{bosfermionHamiltonian} and
\eq{bosCoulombHamiltonian} allow us to write the full quantum
Hamiltonian in terms of the bosonic operators
together with $\oA$.
We have already seen that $\oA$ is an oscillator with frequency
$\omega_0$ given by \Eq{bosmasszero} in each charge sector.
The Hamiltonian is quadratic in the boson fields, and is therefore
equivalent to a set of free oscillators.
It remains to identify the spectrum.  
The oscillator combinations orthogonal to the chiral charge vectors
$q_s$ do not appear in \Eq{bosCoulombHamiltonian}; they remain
free and massless. We choose an orthonormal basis for them,
\beq
\od^I_{\pm,j}
\equiv\sum_{a:s_a=\pm} e^I_{\pm,a}\ob^a_j
\qquad\text{with}\qquad
I = 1, \ldots, N_\pm - 1,
\label{eq:dbasis}
\eeq
where
\beq
e^I_\pm \cdot e^J_\pm =\de^{IJ}
\qquad\text{and}\qquad
e_\pm^I \cdot q_\pm=0.
\eeq
For each $j>0$, the two oscillators in the
charge directions have the Hamiltonian
\beq
\begin{aligned}
\oH^{\mathrm{B}}_j
={}&(\al_j + j)\left[
\oB^\dagger_{+,j}\oB\sub{+,j}
+\oB^\dagger_{-,j}\oB\sub{-,j}
\right]
\\[3pt]
&+\al_j\left[
\oB^\dagger_{+,j}\oB^\dagger_{-,j}
+\oB\sub{-,j}\oB\sub{+,j}
\right],
\end{aligned}
\label{eq:boschargedHamiltonian}
\eeq
where
\beq
\al_j\equiv\frac{g^2q^2}{4\pi j}.
\eeq
A canonical two-mode Bogoliubov transformation is given by
\beq
\begin{aligned}
\oc_{+,j}
&=\cosh\xi_j\,\oB_{+,j}
+\sinh\xi_j\,\oB^\dagger_{-,j},
\\[3pt]
\oc_{-,j}
&=\cosh\xi_j\,\oB_{-,j}
+\sinh\xi_j\,\oB^\dagger_{+,j},
\end{aligned}
\label{eq:bosBogoliubov}
\eeq
with
\beq
\tanh(2\xi_j)=\frac{\al_j}{\al_j + j}.
\eeq
The $\oc_{s,j}$ obey canonical oscillator commutation relations.
In terms of them, \Eq{boschargedHamiltonian} becomes
\beq
\oH^\mathrm{B}_j
=\om_j\left(
\oc^\dagger_{+,j}\oc^{\vphantom{\dagger}}_{+,j}
+\oc^\dagger_{-,j}\oc^{\vphantom{\dagger}}_{-,j}
\right)+\varepsilon_j,
\label{eq:bosdiagonalHamiltonian}
\eeq
with
\beq
\varepsilon_j=\om_j-(\al_j + j)
\qquad\text{and}\qquad
\om_j^2=(\al_j + j)^2-\al_j^2
=j^2+\frac{g^2q^2}{2\pi}
=j^2+\omega_0^2.
\label{eq:bosdispersion}
\eeq
The normalization of the charged oscillators uses anomaly cancellation,
$q_+^2=q_-^2=q^2/2$.  
For large $j$, $\varepsilon_j=O(j^{-3})$, so the shift of the vacuum
energy is finite, showing that diagonalizing the infinitely many oscillator 
pairs does not require  additional UV regularization.

The mass in \Eq{bosdispersion} is equal to the zero-mode mass
\Eq{bosmasszero}, so $\oA$ supplies the zero-momentum member of the
same massive boson whose nonzero modes are generated by the Coulomb
interaction.
These are the modes of an ordinary free massive scalar on $S^1$,
while the orthogonal oscillators describe free massless chiral bosons.
The bosonized theory therefore gives the expected locally relativistic
dynamics, including the effects of the Coulomb vacuum subtraction.
We use this equivalence as our check that the gauge-fixed construction
gives the correct theory, with Lorentz covariance and microcausality
for the local physical observables.

\subsection{Quantum Numbers of the Bosonized States}
All of the bosonic oscillator creation and annihilation operators
preserve the conserved fermion-number charges (see \Eq{oQrdef}):
\beq
[\oQ_r,\ob^a_j]
=[\oQ_r,\oB_{s,j}]
=[\oQ_r,\oc_{s,j}]=0.
\eeq
The gauge zero-mode variables also commute with the conserved
charges. In addition, $\oA$ commutes with the nonzero-momentum
bosonic oscillators:
\begin{align}
[\oQ_r,\oA]
=[\oQ_r,\op{\Pi}]=0
\qquad\text{and}\qquad
[\oA,\ob^a_j]
=[\oA,\oB_{s,j}]
=[\oA,\oc_{s,j}]=0.
\end{align}
Therefore, acting with bosonic raising and lowering operators
does not change the $U(1)$ quantum numbers of a state; these are determined by the charged vacuum on which the bosonic Fock space is constructed.

Another interesting quantum number is fermion parity, defined
in the fermionic description by
\beq
(-1)^{\oF}
\equiv
\exp\left\{
\ii\pi\sum_{a,\ell}\on^a_\ell
\right\}.
\label{eq:fermionparity}
\eeq
This parity is also determined by the conserved charges $C_r$, the eigenvalues of $\hat Q_r$ with $r\cdot q = r\cdot\tilde q = 0$.  
To see
this, two physical charge vectors $f$ and $f'$ with the same $C_r$
differ by
\beq
f'-f=k\frac{\tilde q}{d}
\qquad\text{with}\qquad
d=\gcd\{q_a\}
\qquad\text{and}\qquad
k\in\mathbb Z.
\eeq
Anomaly cancellation, applied to the relatively prime charges
$q_a/d$, implies
\beq
\sum_a\frac{\tilde q_a}{d}
=0\pmod 2,
\eeq
where we used $n^2=n\pmod 2$.  Therefore $f$ and $f'$ have the same
fermion parity.  

\medskip

\noindent\emph{345 Model Example:} In the case of the 345 model, we take $(q_1,q_2,q_3)=(3,4,5)$ and
$(s_1,s_2,s_3)=(+,+,-)$.
There is one conserved fermion-number charge, which we take to be
\beq
C=4f^1-3f^2.
\label{eq:345C}
\eeq
Solving the Gauss law constraint
$3f^1+4f^2+5f^3=0$ gives
\beq
f &=\mat{\hphantom{-}5c+3k\,\\\hphantom{-}5c+4k\,\\-7c-5k\,}
\qquad\text{with}\qquad
c,k\in\ZZ,
\label{eq:345charges}
\eeq
so the charges are 
\begin{subequations}
\beq
C &=5c,
\\[3pt]
\tilde Q
&=\tilde q\cdot f
=70c+50k
=20c\pmod{50}.
\eeq
\end{subequations}
The eigenvalue of the gauge-invariant charge $\op{X}^{\rm inv}$ defined in
\Eq{Xdef} is therefore
\beq
X = e^{4\pi\ii c/5}.
\eeq
Changing $k \to k+1$ shifts $f$ by $\tilde q$, which is equivalent
to an LRGT, so $c=C/5$ labels the inequivalent physical charge sectors.
Their fermion parity is
\beq
(-1)^F=(-1)^c=(-1)^{C/5}.
\label{eq:345parity}
\eeq 

The oscillators along the charged directions are
\beq
\oB_{+,j}=\frac15\left(3\ob^1_j+4\ob^2_j\right)
\qquad\text{and}\qquad
\oB_{-,j}=\ob^3_j.
\eeq
The Coulomb interaction pairs these right- and left-moving charged
towers to form a single ordinary scalar with mass
\beq
\om_0^2=\frac{25g^2}{\pi}.
\eeq
The remaining canonical oscillator, orthogonal to the right-moving
charge vector, is
\beq
\od_{+,j}=\frac15\left(4\ob^1_j-3\ob^2_j\right).
\label{eq:345masslessboson}
\eeq
The massless spectrum therefore consists of exactly
one right-moving chiral boson.

The charge-dependent part of \Eq{boszeropotential} is especially
simple in this example:
\beq
V_f(A)
=25\left(A-\frac75c-k\right)^2+\frac12c^2.
\eeq
The same charge sector has momentum $\frac12c^2$.
Writing $n_0$ for the occupation number of the gauge zero-mode
oscillator, $n_{\pm,j}$ for those of the massive oscillators,
and $n_{d,j}$ for those of the massless right-moving oscillators,
the complete energy and momentum are
\beq
\begin{aligned}
E
&=E_\mathrm{vac}+\frac12c^2+\om_0\gap n_0
+\sum_{j=1}^\infty
\left[\om_j(n_{+,j}+n_{-,j})+j\gap n_{d,j}\right],
\\
P
&=\frac12c^2+\sum_{j=1}^\infty
j\left(n_{+,j}-n_{-,j}+n_{d,j}\right),
\end{aligned}
\label{eq:EandPc}
\eeq
where all occupation numbers are nonnegative integers and
\beq
E_\mathrm{vac}=\frac12\om_0+\sum_{j=1}^\infty\varepsilon_j.
\eeq
This common vacuum energy is finite with our subtraction
prescription.
In particular, odd $c$ gives half-integer momentum and odd
fermion parity, as expected for the antiperiodic fermion
boundary conditions.

\subsection{Matching 't Hooft Anomalies}
We now show that the bosonized theory has the same
't Hooft anomalies as the original fermion theory.  The mixed anomaly
between $C_r$ and the gauged $U(1)$ is trivial to match, and its coefficient is
\beq
\scr{A}_r=r\cdot\tilde q=0
\eeq
for every conserved charge $C_r$.  If this coefficient were nonzero,
$C_r$ would not be conserved in the gauge theory.

However, the conserved global charges can have anomalies among
themselves that must be matched. For any charge $r$, define
\beq
(r_\pm)_a = \begin{cases}
r_a & s_a = \pm, \\
0 & \text{otherwise}.
\end{cases}
\eeq
For a basis $r^\al$ of conserved charges, define
\beq
K^{\al\be} \equiv r^\al_+ \cdot r^\be_+
- r^\al_- \cdot r^\be_-.
\label{eq:globalanomalymatrix}
\eeq
For example, in order to gauge all of these symmetries in the 
fermionic description, we would have to add spectator fermions whose anomaly 
matrix is $-K^{\al\be}$.

To match these anomalies, we need the contribution to the anomaly
from a chiral boson.
The couplings of the massless oscillators in \Eq{dbasis} to background gauge fields for $C_{r^\al}$ are determined by
\beq
R^{\al,I}_\pm
\equiv r^\al_\pm \cdot e^I_\pm =
\sum_{a:s_a=\pm} r^\al_a  e^I_{\pm,a}.
\eeq
A chiral boson labeled by $I$ of chirality $\pm$ contributes
$\pm R^{\al,I}_\pm R^{\be,I}_\pm$ to the anomaly matrix.
The two chiral bosons that form the massive scalar lie in the
$q_+$ and $q_-$ directions.  The conditions
\beq
r^\al\cdot q=0
\qquad\text{and}\qquad
r^\al\cdot\tilde q=0
\eeq
imply
\beq
r^\al_\pm \cdot q_\pm=0.
\label{eq:globalchargeorthogonal}
\eeq
Thus neither chiral component of the massive scalar couples to the
background fields for $C_{r^\al}$, and the massive scalar contributes
zero to $K^{\al\be}$.
The massless chiral bosons' contribution to the anomaly matrix is
given by
\beq
\sum_s \sum_{I=1}^{N_s-1} s R^{\al,I}_s R^{\be,I}_s
= K^{\al\be}
\eeq
as a result of the completeness relations
\beq
\sum_{I=1}^{N_\pm-1} R^{\al,I}_\pm R^{\be,I}_\pm
= \sum_{a:s_a=\pm} r^\al_a r^\be_a.
\eeq
This shows that the massless chiral bosons reproduce the correct
't Hooft anomalies.

The gravitational anomaly is matched similarly.  Every right-moving
chiral fermion or chiral boson contributes $+1$, while every
left-moving one contributes $-1$.  The original fermions therefore
have anomaly coefficient $N_+-N_-$,
while the massless spectrum contains $N_+-1$
right-moving and $N_--1$ left-moving chiral bosons, 
which have anomaly coefficient
$(N_+-1)-(N_--1) =N_+-N_-$.

\section{Truncation and Numerical Results}
\label{sec:Numerics}
In this section, we turn to the numerical evaluation of the Hamiltonian framework developed above. 
We will use Hamiltonian Truncation (HT)~\cite{FitzpatrickKatz2022} to compute the spectrum of the vector-like Schwinger model and the chiral 345 model.

The Hamiltonian truncation approach can be summarized as follows.
 Take the continuum Hamiltonian for a given quantum field theory, defined as a relevant deformation of a solvable UV theory. In our construction, the first step is to put the theory in finite volume so that the spectrum becomes discrete. One then chooses an exactly solvable Hamiltonian $\oH_0$, which is often taken to be the Hamiltonian of the free particles in the theory. This free Hamiltonian is then used to define a basis of states, typically the Fock basis. The full Hamiltonian of the theory is then expressed as $\oH = \oH_0 + \oV$, and we will use the obvious terminology and refer to $\oV$ as the `potential.'  One then introduces an `energy cutoff' $\Emax$, and defines a truncated basis of states $\{\ket{n}\}$ such that $\oH_0 \ket{n} = E_{n} \ket{n}$ with $E_{n} \leq \Emax$. One can explicitly compute the matrix elements of the full Hamiltonian in this basis $\bra{m} \oH \ket{n}$.

This can be thought of as defining the theory with a UV cutoff $E_\mathrm{max}$.
 In general, the theory requires counterterms in order for the limit $E_\mathrm{max} \to \infty$ to exist.
 The Hamiltonian truncation UV cutoff is subtle because it is nonlocal and not Lorentz invariant.
In simple low-dimensional examples such as the theory we are considering, no counterterms are required for convergence.
 In fact, the arguments of \cite{Cohen:2021erm} predict that the low-lying energy differences should converge as $1/E_\mathrm{max}^3$.
 We can therefore numerically compute the spectrum directly from the truncated Hamiltonian, and the results are consistent with this prediction.
 As emphasized above, our approach is completely gauge invariant and has exact chiral symmetry.
 
 Hamiltonian truncation has been developed and applied across a wide range of theories and observables~\cite{Brooks:1983sb,Pauli:1985ps,Yurov:1989yu,Yurov:1991my,Katz:2013qua,Hogervorst:2014rta,Rychkov:2014eea,Rychkov:2015vap,Bajnok:2015bgw,Katz:2016hxp,Rakovszky:2016ugs,James:2017cpc,Anand:2017yij,Hogervorst:2018otc,Fitzpatrick:2019cif,Elias-Miro:2020qwz,Anand:2020gnn,Anand:2020qnp,Hogervorst:2021spa,Chen:2021bmm,EliasMiro:2021aof,Emonts:2022vim,Chen:2022zms,Delacretaz:2022ojg,Henning:2022xlj,Fitzpatrick:2023mbt,Fitzpatrick:2023aqm,Schmoll:2023eez,Ingoldby:2024fcy,Fitzpatrick:2024rks,Ingoldby:2025bdb,Basak:2026meu,Houtz:2026wra}. 
The method can be understood in the language of Effective Theory~\cite{Cohen:2021erm,Demiray:2025zqh,Maestri:2026hqb, Delouche:2026cqe}
to compute counterterms and subleading corrections in an expansion in the cutoff
(see also \cite{Feverati:2006ni,Giokas:2011ix,Lencses:2015bpa,Elias-Miro:2015bqk,Elias-Miro:2017tup,Rutter:2018aog,Fitzpatrick:2018ttk,EliasMiro:2022pua,Chen:2023glf,Lajer:2023unt,Delouche:2023wsl,Delouche:2024yuo,Li:2026dyb}).
Hamiltonian truncation has previously been applied to two-dimensional gauge theories using both equal-time and lightcone constructions~\cite{Katz:2014uoa, Anand:2021qnd,Houtz:2025lbv}, and to special limits of three-dimensional gauge theories~\cite{Delacretaz:2018xbn,Fitzpatrick:2025hqk}.

For the present theories, we choose the decomposition
\begin{align}
    \oH_0 &=\oH_{\rm 0, ferm} + \oH_\mathrm{zm}
    \qquad\text{and}\qquad
    \oV = \oH_\mathrm{Coul},
\end{align}
where $\oH_{\rm 0, ferm}$ is the free-fermion Hamiltonian from \Eq{H0ferm}, and $\oH_\mathrm{Coul}$ is given in \Eq{HCoulQ}.
As above, we set $R = 1$, where $R$ is the radius of the spatial $S^1$.
The gauge zero-mode contribution to $\oH_0$ is
\begin{align}
    \oH_\mathrm{zm}
    = \frac{g^2}{4\pi} \hat\Pi^2 
    - \hat A\,\hat{\tilde Q}
    + \frac{1}{2}q^2\hat A^2,
    \label{eq:zm}
\end{align}
where we have dropped the $\th$ term (see \sec{GaugeInvStates}).
Here $\op{\tilde{Q}}$ is the normal-ordered axial charge, which is not invariant under LRGT, but $\oH_0$ is invariant.

We now show that $\oH_0$ can be diagonalized exactly despite the coupling of the fermions to $\oA$.
The free-fermion Fock states are simultaneous eigenstates of $\oH_{0,\mathrm{ferm}}$ and $\op{\tilde Q}$. 
Because $[\oH,\op{\tilde Q}]=0$, the Hilbert space decomposes into independent sectors labeled by the eigenvalues $\tilde Q$. Within each such sector, the zero-mode contribution reduces to a shifted harmonic oscillator, which can be brought to the form
\begin{align}
    \oH_\mathrm{zm}
    = \omega^{\vphantom{\dagger}}_0\gap\oc^\dagger_0\gap \oc^{\vphantom{\dagger}}_0
    - \frac{\tilde Q^2}{2q^2},
    \label{eq:hzmfinal}
\end{align}
where $\oc^\dagger_0$ and $\oc^{\vphantom{\dagger}}_0$ are the creation and annihilation operators for the zero mode, and $\omega_0$ is defined in Eq.~(\ref{eq:bosmasszero}). 

The eigenstates of $\oH_0$ can therefore be constructed
from the fermionic Fock states and the zero-mode
oscillator eigenstates, providing a basis for
Hamiltonian truncation in each fixed-$\tilde Q$
sector. The $E_{\max}$ cutoff is imposed on the fermionic Fock states, while the zero-mode oscillator is treated separately. The constant contribution
$-\tilde Q^2/(2q^2)$ must be retained when
comparing energies between different sectors. We neglect the conventional vacuum energy contribution of $\omega_0/2$ and other sector-independent additive constants
in Eq.~(\ref{eq:hzmfinal}), which have no effect on energy differences.

We also restrict the truncated Hilbert space to eigenstates of $\oH_0$ with specified eigenvalues of the momentum $\oP$ and the gauge, axial, and global $U(1)$ charges $\oQ_r$. These operators each commute with $\oH_0$ and $\oV$, and can be expressed in terms of the fermionic mode occupation numbers as
\begin{align}
    \oP &= \sum_{a,\ell}s_a|\ell|\on^a_\ell 
    \qquad\text{and}\qquad
    \oQ_r = \sum_{a,\ell}s_a r_a\operatorname{sgn}(\ell)\,\on^a_\ell,
\end{align}
using $\on^a_\ell \equiv \oa^{a\,\dagger}_\ell \oa^a_\ell$ to denote the number operator for mode $\ell$ of species $a$. The sign of the momentum carried by each mode is determined solely by its species chirality $s_a=\pm 1$, whereas the sign of its contribution to the $U(1)$ charges also depends on the sign of $\ell$. Restricting to these sectors also constrains the total fermion number to be either even or odd for all basis states.

Since fermionic creation operators anticommute, their ordering must be specified to define each Fock basis state (\Eq{Fockstate}) unambiguously. In our numerical Hamiltonian truncation calculations, we adopt the following ordering convention to ensure consistent signs in the Hamiltonian matrix elements, proceeding from left to right: (i) increasing gauge charge $q_a$; (ii) positive $\ell$ before negative $\ell$; (iii) positive chirality before negative chirality; and (iv) increasing $|\ell|$. Each criterion is applied only when the preceding criteria do not distinguish the modes.

Since $\oV$ is independent of the gauge zero-mode oscillator, the zero-mode occupation number is conserved. We therefore restrict our numerical calculations to the zero-mode ground state, with excited states recovered by adding integer multiples of $\omega_0$ to the corresponding energies. All truncated Hilbert space dimensions quoted in the tables below therefore exclude states with nonzero zero-mode occupation number.

When computing the matrix elements of $\oV$ in the truncated basis, we use the following expressions for species currents in terms of the operators $\oa^{a\vphantom{\dagger}}_\ell$ and $\oa^{a\gap\dagger}_\ell$. For momentum index $j>0$ and a species $a$ with chirality $s_a$, we use
\begin{align}
    \orho^a_{s_a j}
    &= \sum_{\ell>0}
    \left(
        \hat a^{a\gap\dagger}_{s_a\ell} \hat a^{a\vphantom{\dagger}}_{s_a(\ell+j)}
        - \hat a^{a\gap\dagger}_{-s_a\ell} \hat a^{a\vphantom{\dagger}}_{-s_a(\ell+j)}
    \right)
    + \sum_{0<\ell<j}
        \hat a^{a}_{-s_a\ell} \hat a^{a}_{s_a(j-\ell)},
    \label{eq:species-current}
\end{align}
while the opposite momentum equivalents can be found using $\orho^a_{-s_a j}= \left(\orho^a_{s_a j}\right)^\dagger$. Because the Hamiltonian matrix elements are evaluated exactly within the retained subspace, the raw eigenvalues satisfy the variational bound and approach the exact finite-volume energies from above as $E_{\rm max}$ is increased, a behavior we verify as a consistency check.

We focus on excitation energies, {\it i.e.}~energy level differences of the form $\Delta E_n\equiv E_n -E_0$. These differences are unaffected by any scheme-dependent offsets in the vacuum energy and can be directly compared with the exact results known from bosonization; they should correspond to sums of single-particle energies of free fermions and scalars. We emphasize that bosonization is only used here as a verification that our numerical results are asymptoting to the correct values. All numerical evaluation is done using the description of the theory in terms of fermions and the gauge zero mode.

The convergence of the spectrum is predicted by the effective Hamiltonian approach of \cite{Cohen:2021erm}.
The allowed Hamiltonian counterterms consist of all terms of the form $\int \text{d}x\, \op{\scr{O}}_\mathrm{ct}$, where the counterterm operators are allowed to be nonlocal through their dependence on $H_0$, which measures the Hamiltonian truncation cutoff.
The coefficients then depend on $E_\mathrm{max}$ as determined by dimensional analysis.
The free theory has only a vacuum counterterm proportional to the identity operator.
The nontrivial counterterms appear at second order in perturbation theory, so they must be proportional to $g^4$, which has mass dimension 4.
The lowest-order counterterms are therefore
\beq
\oH_\mathrm{ct} &= \frac{g^4}{E_\mathrm{max}^2} \int \text{d}x \bigg[ c_1 
+ c_2 \frac{\oH_{0}}{E_\mathrm{max}} + O(1/E_\mathrm{max}^2) \bigg]
+ O(g^6/E_\mathrm{max}^4),
\label{eq:scaling}
\eeq
where the $c_i$ are dimensionless.
The constant vacuum energy subtraction does not affect energy differences, so this predicts that the error in the energy differences is controlled by the nonlocal $\oH_0$ term.
We therefore expect the errors for these  to scale as $1/E_\mathrm{max}^3$.

 The Hamiltonian is constructed separately in each symmetry sector and diagonalized as a finite sparse matrix. For both the Schwinger and $345$ models, the numerical spectrum can be compared directly with the exact result obtained from bosonization. We first consider the vector-like Schwinger model as a simple test of the numerical calculation before turning to the chiral theory.

\subsection{The Schwinger Model}

The vector-like Schwinger model consists of a single massless Dirac fermion,
corresponding to two Weyl fermions with $(q_1,q_2)=(1,1)$ and
$(s_1,s_2)=(+,-)$. The gauge and axial charges are
$Q=f^1+f^2$ and $\tilde Q=f^1-f^2$, respectively.

\begin{table}[h!]
    \centering
    \begin{tabular}{c c}
        \hline
        $E_{\max}$ & Truncated basis dimension \\
        \hline
        8 & 40 \\
        12 & 210 \\
        16 & 919 \\
        20 & 3,583 \\
        24 & 12,648 \\
        \hline
    \end{tabular}
    \caption{Truncated basis dimensions used for calculating the spectrum of the massless Schwinger model for increasing values of the cutoff $E_{\max}$.}
    \label{tab:basis-dimension}
\end{table}

\begin{figure}[b!]
   \centering
   \begin{minipage}{.9\textwidth}
   \centering
   \begin{minipage}[t]{0.48\textwidth}
\vspace{0pt}
\centering
\includegraphics[width=\linewidth]{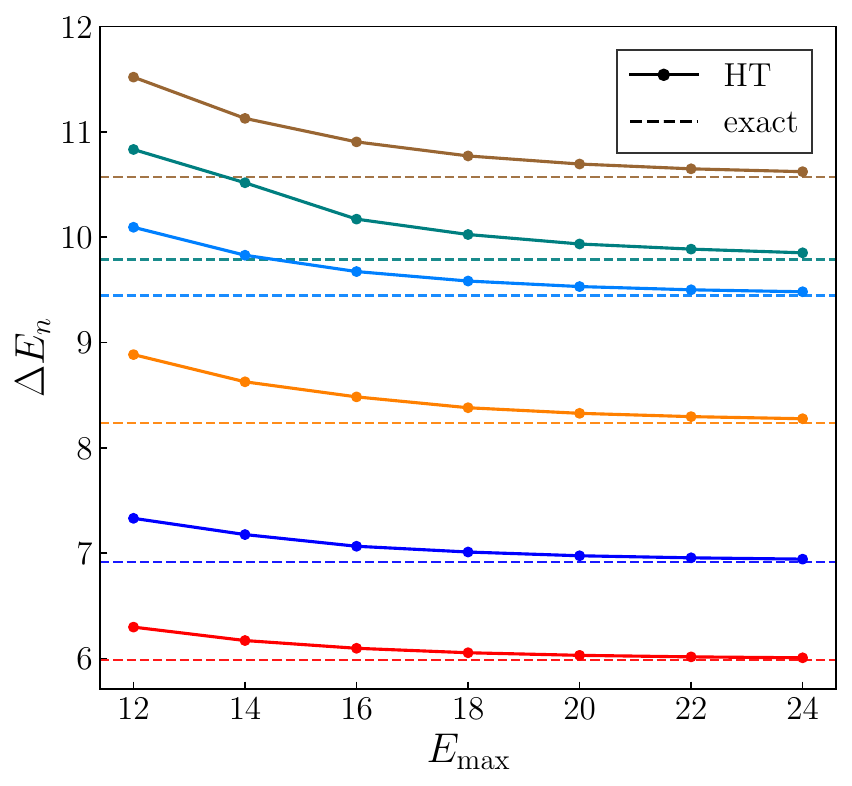}
\end{minipage}
\hspace{1pt}
\begin{minipage}[t]{0.48\textwidth}
\vspace{0pt}
\centering
\includegraphics[width=\linewidth]{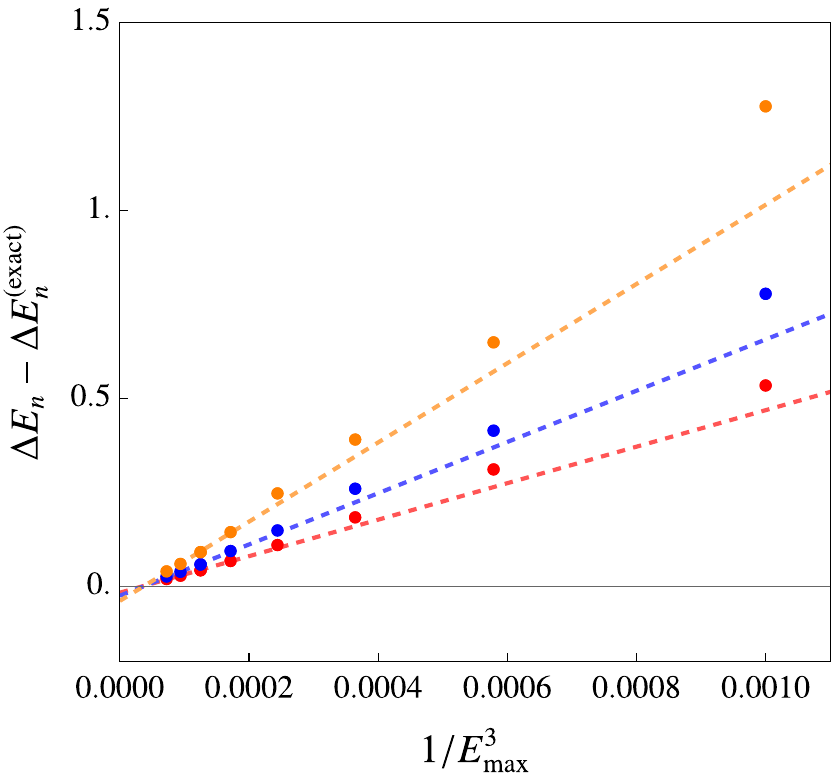}
\end{minipage}
   \caption{The left plot shows the six lowest distinct (ignoring degeneracies) energy gaps $\Delta E_n$ of the Schwinger model in the $Q=\tilde Q =P=0$ sector at $g=5$, as a function of $E_{\max}$. The dashed lines show the exact results from bosonization, which are (from lowest to highest): $2\omega_1$, $2\omega_2$, $2\omega_3$, $2\omega_1+\omega_2$, $2\omega_4$, and $\omega_1+\omega_2+\omega_3$. On the right, we show the difference between the lowest energy gaps $\Delta E_n$ and the exact values from bosonization $\Delta E_n^{(\rm exact)}$ of the Schwinger model given by \Eq{Schwingerexact} as a function of $1/E_{\rm max}^3$ at $g=5$. The three gaps in the right plot correspond to exact energies of $2\omega_1$, $2\omega_2$, and $2\omega_3$, in order from smallest to largest $\Delta E_n$. This ordering is preserved for the quantity $\Delta E_n - \Delta E_n^{\rm (exact)}$ plotted on the right. The dashed lines are linear fits in $1/E_{\max}^3$ using the five largest values of $E_{\max}$. We use $R=1$ in both plots. 
   }
   \label{fig:schwinger-convergence}
   \end{minipage}
\end{figure}

As reviewed in the bosonization section, the exact finite-volume spectrum of the Schwinger model is that of a free massive scalar with squared mass given by
\begin{align}
\omega_0^2=\frac{g^2}{\pi}.
\end{align}
This gives a direct comparison with the truncated fermionic calculation. We work in the neutral, zero-momentum sector with $Q=\tilde Q=P=0$. The exact energy gaps, with the zero-mode oscillator in its ground state, are
\begin{align}
\Delta E^{(\rm exact)}_n=\sum_{j\neq 0}n_j\gap\omega_j \qquad \text{with}\qquad
\omega_j=\sqrt{j^2+\frac{g^2}{\pi}},
\label{eq:Schwingerexact}
\end{align}
where $n_j$ are nonnegative integer occupation numbers of modes with momentum $j$ satisfying $\sum_{j\neq 0}j n_j=0$ in the $P=0$ sector. As indicated in Table~\ref{tab:basis-dimension}, for the largest cutoff used,
$E_{\rm max}=24$, the truncated fermionic basis contains $\sim 13,\!000$ states.

In Fig.~\ref{fig:schwinger-convergence}, the left plot shows the lowest energy gaps $\Delta E_n = E_n - E_0$ as a function of $E_{\rm max}$ for $g = 5$. The gaps converge smoothly to the exact spectrum given by \Eq{Schwingerexact} as the cutoff is increased. The right plot shows the difference between the lowest energy gaps and their exact bosonized values at $g=5$ as a function of $1/E_{\max}^3$. The approximately linear behavior is consistent with the expected scaling from~\Eq{scaling}, confirming the prediction from the effective Hamiltonian.

\begin{figure}[h!]
   \centering
\includegraphics[width=.6
\linewidth, clip=true, trim=.1cm .1cm 0cm .2cm]{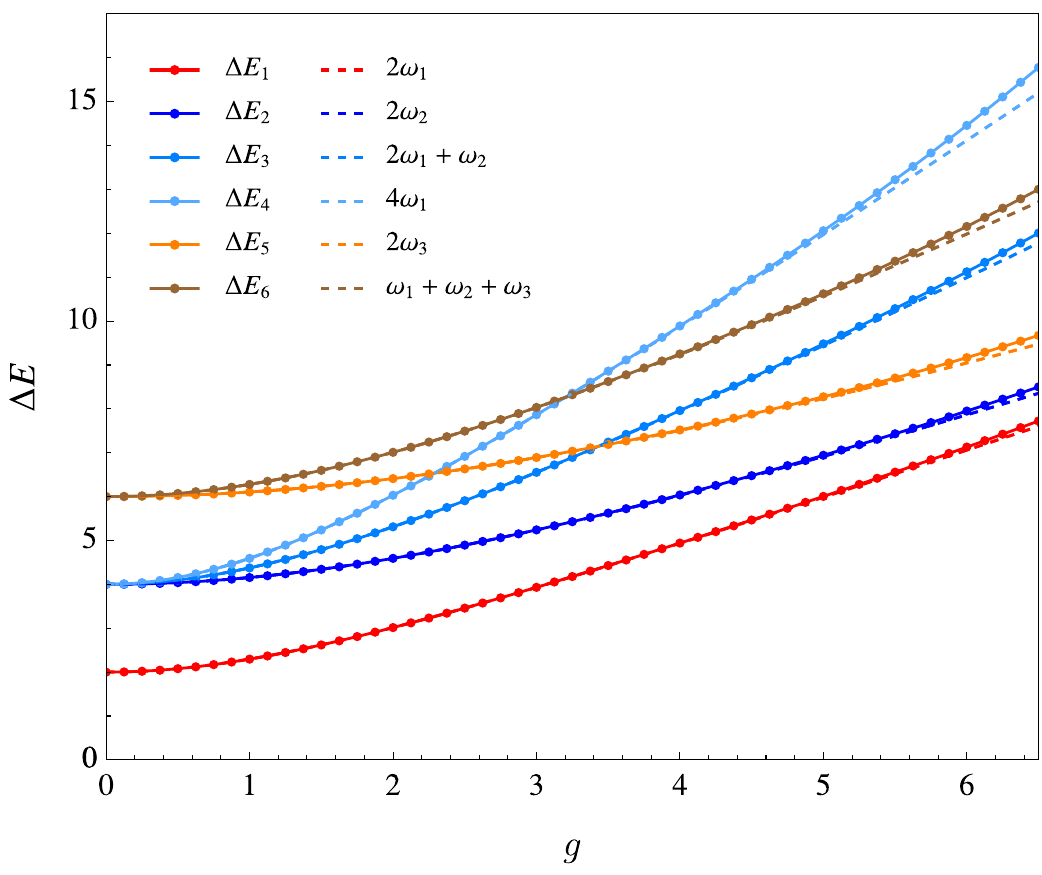}
   \caption{
   Low-lying energy gaps of the Schwinger model as a function of the gauge coupling at $E_{\rm max}=24$. The markers show the Hamiltonian truncation results, with solid lines providing a linear interpolation between points to guide the eye. The dashed lines show the exact bosonized spectrum. Gauge zero-mode oscillator excitations are omitted.
   }
   \label{fig:schwinger-spectrum}
\end{figure}

Finally, in Fig.~\ref{fig:schwinger-spectrum} we fix $E_{\rm max}=24$ and vary the gauge coupling. When $g$ is small (in units of the radius, $R$, which we set to 1), the numerical spectrum matches the exact bosonized result with very high accuracy. Small deviations between the numerical and bosonized energy levels become visible at larger couplings, consistent with the predicted scaling in $g$ from \Eq{scaling}. As the coupling is increased, all the energy levels increase, but level crossings occur, as permitted in a theory of free massive bosons.\footnote{Note that the level ordering in Fig.~\ref{fig:schwinger-convergence} is therefore not the ordering of the theory when $g\ll 1$.}
 In particular, levels containing larger numbers of bosons carrying less momentum are lifted above those containing fewer bosons with more momentum. We show only the six energy gaps in the $P=0$ sector which have the lowest energies at zero coupling.

\subsection{The {\normalfont 345} Model}
We now turn to the chiral $345$ model. The model includes three Weyl fermions with $(q_1,q_2,q_3)=(3,4,5)$ and $(s_1,s_2,s_3)=(+,+,-)$. The gauge charge is $Q=3f^1+4f^2+5f^3$, the axial charge is $\tilde Q=3f^1+4f^2-5f^3$, and the conserved global $U(1)$ charge is $C=4f^1-3f^2=5c$. The bosonized spectrum derived in \sec{Bosonization}
contains a massive scalar with
\begin{align}
\omega_0^2=\frac{25g^2}{\pi},
\label{eq:345mass}
\end{align}
and a massless right-moving chiral boson. On the circle, the chiral boson can equivalently be described by a free right-moving chiral fermion with antiperiodic boundary conditions, as shown in \sec{Bosonization}.

We note that adding an extra spectator fermion with zero gauge charge and negative chirality would convert the $345$ model to the widely studied $3450$ model, reviewed in Refs.~\cite{Wang:2022ucy,Mouland:2025dct}. However we omit this spectator in our study, since omitting it reduces the size of the Hilbert space needed for numerical calculations. Both models are free of gauge anomalies, but the unequal numbers of positive and negative chirality fermions in the $345$ model give it a gravitational anomaly that is absent in the $3450$ model. This creates no obstruction to applying Hamiltonian truncation however, which requires neither coupling to a gravitational background nor the introduction of a spatial lattice.

\begin{table}[b]
    \centering
    \begin{tabular}{c c}
        \hline
        $E_{\max}$ & Truncated basis dimension \\
        \hline
        8 & 143 \\
        12 & 1,110 \\
        16 & 6,830 \\
        20 & 36,032 \\
        24 & 167,849 \\
        28 & 706,394 \\
        32 & 2,747,532 \\
        \hline
    \end{tabular}
    \caption{Truncated basis dimensions obtained for calculating the spectrum of the 345 model in the $P=0$, $c=0$ symmetry subsector for increasing values of the cutoff $E_{\max}$.}
    \label{tab:basis-dimension345}
\end{table}

Using the free boson and chiral fermion description, the exact energy gaps are (again with the gauge zero mode in its ground state):
\begin{align}
\Delta E_n^{(\rm exact)} = \sum_{j\neq 0} n_j \omega_j + \sum_{\ell >0} \left(n_{\ell}^{(+1)} \bar{\omega}_\ell^{(+1)}+n_{\ell}^{(-1)} \bar{\omega}_\ell^{(-1)}\right),
\end{align}
where 
\begin{align}
\omega_j = \sqrt{j^2 + \frac{25 g^2}{\pi}} 
\qquad \text{and} \qquad
\bar{\omega}^{(\pm 1)}_\ell = \ell \quad\text{with}\quad \ell  = 1/2,\ 3/2,\  \cdots\,.
\label{eq:bosonized345}
\end{align}
Here $\omega_j$ and $\bar{\omega}_\ell^{(\pm1)}$ are the single-particle energies of the massive scalar and massless right-moving chiral fermions, respectively. The bosonic occupation numbers $n_j$ are nonnegative integers, while the fermionic ones $n_\ell^{(\pm1)}\in\{0,1\}$ have superscripts $(\pm1)$ indicating the individual fermion contributions to the total charge $c=C/5$.

\begin{figure}[b!]
   \centering
    \includegraphics[width=0.45\linewidth]{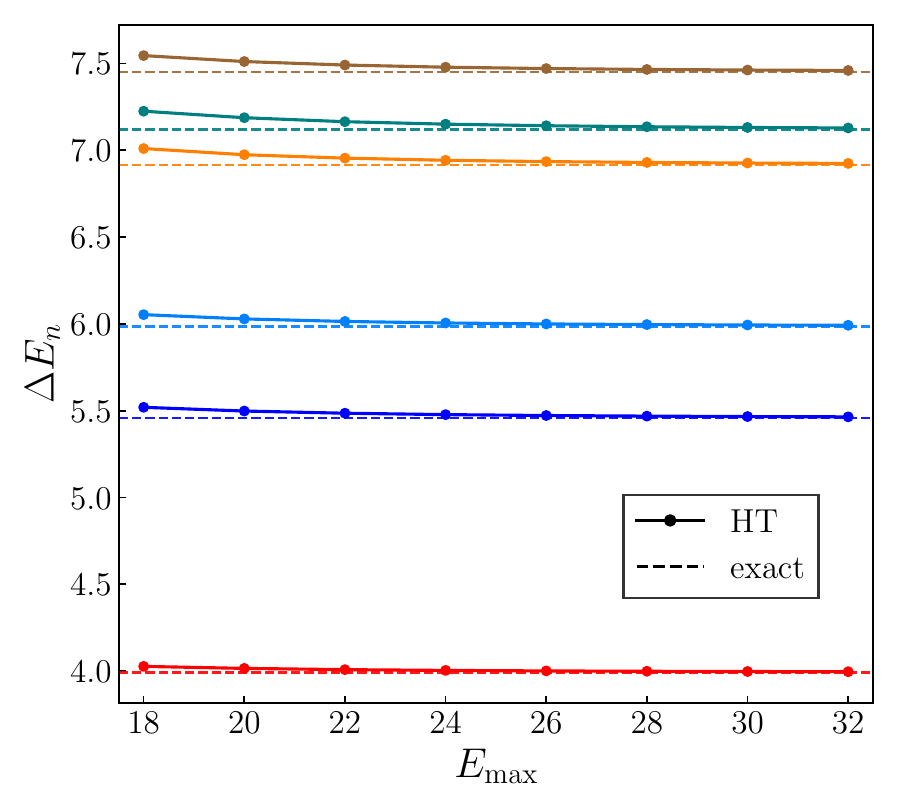}
   \includegraphics[width=0.45\linewidth]{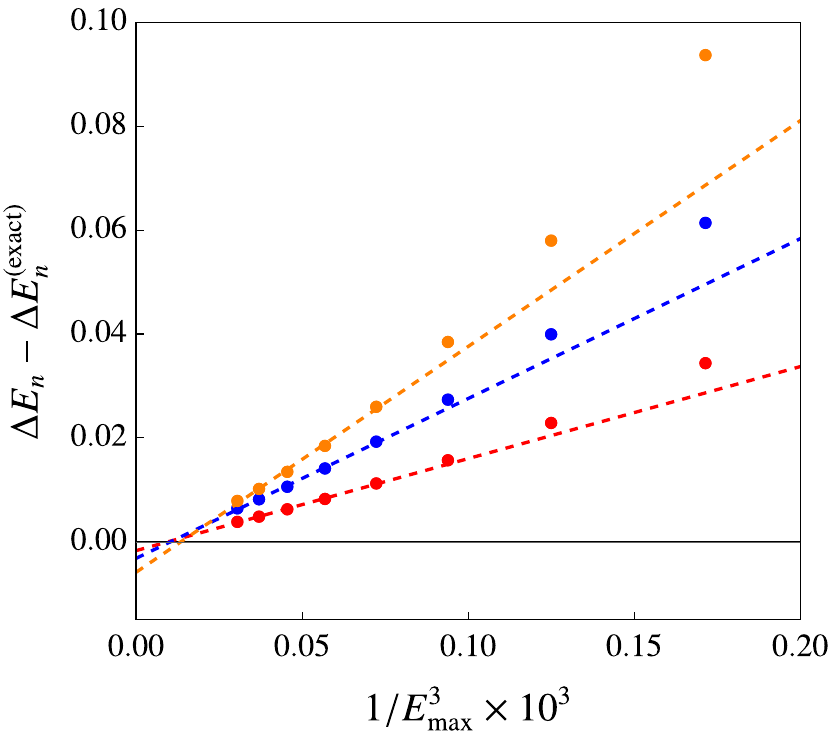}
   \caption{The left plot shows the six lowest distinct energy gaps of the 345 model in the $c=0$, $P=0$ sector at $g=1$ as a function of the truncation cutoff $E_{\max}$. Solid lines with markers show the Hamiltonian truncation results, while dashed lines show the exact bosonized energies.
    The right plot shows deviations of selected lowest energy gaps from their exact bosonized values, $\Delta E_n-\Delta E_n^{(\mathrm{exact})}$, for the 345 model in the $c=0$, $P=0$ sector at $g=1$, as a function of $1/E_{\max}^3$. The dashed lines are linear fits in $1/E_{\max}^3$ using the five largest values of $E_{\max}$. 
   }
   \label{fig:345-convergence}
\end{figure}

We first examine the charge-neutral, $P=0$ sector at $g=1$, chosen so that the massive scalar has the same mass as in the Schwinger model at $g=5$. In the 345 model, this corresponds to the sector for which the gauge charge $Q$, the axial charge $\tilde Q$, and the global $U(1)$ charge $c\equiv C/5$ vanish. In this sector, the occupation numbers satisfy
\begin{align}
\sum_{\ell>0}\left(n_\ell^{(+1)}-n_\ell^{(-1)}\right)=0
\qquad\text{and}\qquad
\sum_{j\ne0}j n_j+
\sum_{\ell>0}\ell\left(n_\ell^{(+1)}+n_\ell^{(-1)}\right)=0.
\end{align}
We note that the right-moving chiral fermions only carry positive momentum. Therefore scalars carrying negative momentum need to be included in every excited state in this sector so that $P=0$ overall. At $E_{\rm max}=32$, the $c=P=0$ truncated basis contains $\sim 2.7\times 10^6$ states; the basis sizes for smaller values of $E_{\rm max}$ are provided in Table~\ref{tab:basis-dimension345}.

In the left panel of Fig.~\ref{fig:345-convergence}, we fix $g=1$ and show the lowest energy gaps as a function of $E_{\max}$. They converge toward the exactly known levels in the bosonized description as the cutoff is increased. In the notation of \Eq{bosonized345}, these levels, in order of increasing energy, are $\omega_{-1}+\bar{\omega}^{(+1)}_{1/2}+\bar{\omega}^{(-1)}_{1/2}$, followed by $\omega_{-2}+\bar{\omega}^{(+1)}_{1/2}+\bar{\omega}^{(-1)}_{3/2}$ and its degenerate charge conjugate, $\omega_1+\omega_{-1}$, and $\omega_2+\omega_{-2}$. The next level is $\omega_{-3}+\bar{\omega}^{(+1)}_{3/2}+\bar{\omega}^{(-1)}_{3/2}$, which is triply degenerate with $\omega_{-3}+\bar{\omega}^{(+1)}_{1/2}+\bar{\omega}^{(-1)}_{5/2}$ and its charge conjugate. Finally, we have $\omega_{-2}+\omega_{1}+\bar{\omega}^{(+1)}_{1/2}+\bar{\omega}^{(-1)}_{1/2}$.

\begin{figure}[b!]
   \centering

   \includegraphics[width=.7\linewidth, clip=true, trim=.1cm .2cm .7cm .1cm]{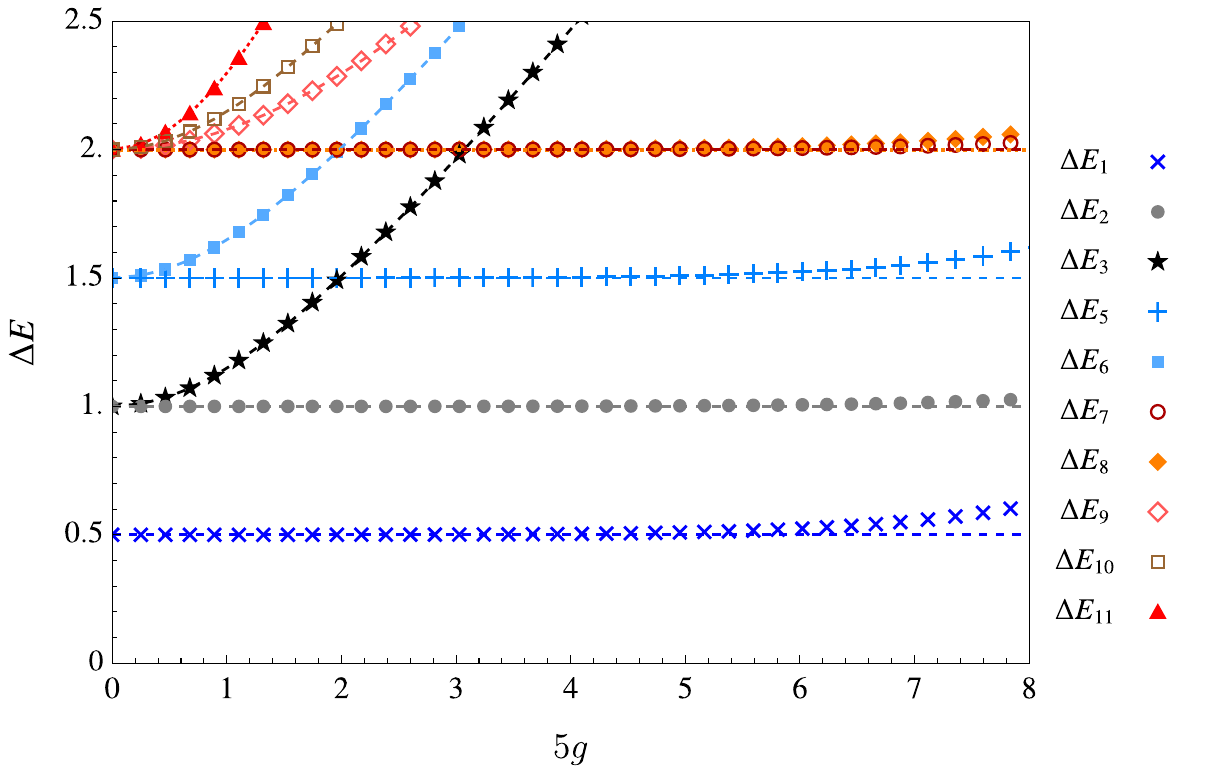}
   \caption{ 
   Selected low-lying energy gaps of the $345$ model in several
   charge and momentum sectors as a function of the gauge coupling at
   $E_{\rm max}=32$, plotted against $5g$, the Schwinger-model coupling with the same scalar mass, given in ~\Eq{345mass}. The markers show the HT results while the dashed lines show the exact bosonized spectrum. The global charge and momentum quantum numbers for the states corresponding to the plotted $\Delta E_n$ are shown in Table~\ref{tab:345speclabels}, along with the occupation numbers in the free boson and chiral fermion description. Only one of the two states labeled $\Delta E_{3,4}$ in Table~\ref{tab:345speclabels} is plotted, since their exact energies are degenerate.
   }
   \label{fig:345-sectors}
\end{figure}

In the right panel, we plot the deviations of three of these energy gaps from their bosonized values against $1/E_{\max}^3$. These are the first, second, and fourth gaps listed above, shown in the same colors as in the left panel. As in the Schwinger model, this dependence becomes linear for large values of $E_{\max}$, consistent with the expected asymptotic scaling in~\Eq{scaling}. Deviations from linearity at smaller values of $E_{\max}$ are also visible in the figure.

We next compare selected low-lying states in several charge and momentum sectors. To resolve the massless fermion sector, we must also consider states with $P\neq 0$, since chiral fermions carry momentum of a fixed sign. Within each fixed sector, the states are followed by their eigenvector overlaps as $g$ is varied, allowing the same state to be tracked through level crossings. Figure~\ref{fig:345-sectors} shows the resulting spectrum at $E_{\rm max} = 32$ up to $g=8/5$. The global charges, momenta and occupation numbers in the free boson and chiral fermion description of all the states plotted are recorded in Table~\ref{tab:345speclabels}.

\begin{table}[b!]
\centering
\renewcommand{\arraystretch}{1.3}
\begin{tabular}{c|ccc@{${}{}$}cc}
$\Delta E_n$ & $c$ & $P$ & \multicolumn{2}{c}{State label} & $\Delta E_n^{\rm (exact)}$\\
\hline

$\Delta E_1$ & $1$ & $\tfrac12$ &
$\bar\omega_{1/2}^{(+1)}$ && $\tfrac12$ \\

 $\Delta E_2$ & $0$ & $1$ &
$\bar\omega_{1/2}^{(+1)} + \bar\omega_{1/2}^{(-1)}$ && $1$ \\

$\Delta E_{3,4}$ & $0$ & $\pm 1$ &
$\omega_{\pm1}$ && $\sqrt{1 + 25g^2/\pi}$ 
\\

$\Delta E_5$& $1$ & $\tfrac32$ &
$\bar\omega_{3/2}^{(+1)}$ && $\tfrac32$ 
\\

$\Delta E_6$ &$1$ & $-\tfrac12$ &
$\bar\omega_{1/2}^{(+1)} + \omega_{-1}$ &&
$\tfrac12 + \sqrt{1 + 25g^2/\pi}$   \\

$\Delta E_7$ & $0$ & $2$ &
$\bar\omega_{1/2}^{(\pm1)} + \bar\omega_{3/2}^{(\mp1)}$ && $2$ \\

$\Delta E_8$ & $2$ & $2$ &
$\bar\omega_{1/2}^{(+1)} + \bar\omega_{3/2}^{(+1)}$ && $2$ \\

$\Delta E_9$ & $0$ & $2$ &
$\omega_2$ && $\sqrt{4 + 25g^2/\pi}$ \\

$\Delta E_{10}$& $0$ & $0$ &
$\bar\omega_{1/2}^{(+1)} + \bar\omega_{1/2}^{(-1)} + \omega_{-1}$ &&
$1 + \sqrt{1 + 25g^2/\pi}$ \\

$\Delta E_{11}$ &$0$ & $0$ &
$\omega_1+\omega_{-1}$ && $2\sqrt{1 + 25g^2/\pi}$
 \\
\end{tabular}
\caption{Charge sector, momentum, and exact energy gap for each state
shown in Fig.~\ref{fig:345-sectors}. Here
$\omega_j=\sqrt{j^2+\omega_0^2}$, with $\omega_0^2=25g^2/\pi$,
denotes the energy of a massive boson, while
$\bar\omega_\ell^{(\pm1)}=\ell$, with
$\ell=\frac12,\frac32,\ldots$, denotes the energy of a massless
right-moving fermion with charge $\pm1$ under $c$.
The column labeled $c$ gives the total charge of the state.
The $\Delta E_{3,4}$ row represents two degenerate states with
$P=+1$ and $P=-1$.
The $\Delta E_7$ row likewise represents two degenerate states,
corresponding to the two assignments of the charge superscripts
$(+1,-1)$ and $(-1,+1)$.
The $\Delta E_8$ row represents a single state in the $c=2$ sector.}
\label{tab:345speclabels}
\end{table}

The Hamiltonian truncation method resolves both the massive and massless sectors. In the exact spectrum, the gaps of states containing only massless fermions are independent of $g$, whereas those containing massive scalars increase with $g$. The numerical levels closely follow the exact bosonized curves. As in the Schwinger model, deviations are too small to be visible at weak coupling, but become apparent at the largest values of $g$ evaluated here.

\section{Conclusions}
In this paper, we presented a Hamiltonian framework for Abelian chiral gauge theories in 1+1 dimensions. Using Hamiltonian truncation, we computed the non-perturbative spectra of the Schwinger and 345 models directly in the fermionic formulation and demonstrated agreement with the exact results from bosonization.
This is a significant step  
in the non-perturbative simulation of chiral gauge theories.

Along the way, we filled in some gaps in the literature regarding structural aspects of two-dimensional Abelian chiral gauge theories, including constructing the physical states in the gauge-fixed theory, identifying the physical $\bar\theta$ parameter, developing the bosonization solution explicitly, and deriving the physical charge sectors.
We also showed the nontrivial role played by the anomaly cancellation condition on the consistency of the theory.

It remains to be seen whether the methods we have developed can be extended to non-Abelian gauge theories and/or theories in higher dimensions.
Hamiltonian truncation becomes more challenging in higher dimensions because of the presence of additional nonlocal counterterms.
However, we emphasize that a completely gauge-fixed Hamiltonian formulation such as the one we have presented may be useful for other numerical approaches, for example tensor networks, or quantum algorithms.
This work already demonstrates that this approach can be used to extract non-perturbative numerical data about foundational questions in quantum field theory that are difficult to access using other techniques.

\section*{Acknowledgments}
We thank 
M. Anber,
P. Bedaque,
S. Sen, S. J. Rey and D.~Tong
for useful discussions. We acknowledge the use of AI tools for coding, checking calculations and to assist with the writing; the authors are fully responsible for the results and content of the manuscript. 
TC is supported by the U.S.~Department of Energy Grant No.~DE-SC0011640. RH is supported by the Institute for Fundamental Theory at the University of Florida.

\clearpage
\addcontentsline{toc}{section}{Appendix A: Conventions}
\section*{Appendix: Conventions}
We use the metric conventions
\beq
g_{\mu\nu} = \mat{1 & 0 \\ 0 & -1},
\qquad
\ep^{\mu\nu} = \mat{0 & 1 \\ -1 & 0}.
\eeq
Our Dirac matrices are given by
\beq
\ga^0 = \mat{0 & 1 \\ 1 & 0},
\qquad
\ga^1 = \mat{0 & 1 \\ -1 & 0}.
\label{eq:Diracmat}
\eeq

For one Dirac fermion, the Weyl species are $\psi^1$ and $\psi^2$, with
$s_1=+1$ and $s_2=-1$, and the spinor is
\beq
\Psi = \mat{\psi^2 \\ \psi^1}.
\eeq
The massless Dirac Lagrangian is
\beq
\scr{L}_\text{Dirac} &= \bar{\Psi} \ii \sla\d \Psi 
= \sum_{a=1}^2\psi^{a\dagger}\ii\d_{s_a}\psi^a
\qquad\text{with}\qquad
\d_\pm = \d_0 \pm \d_1.
\label{eq:DiracLconv}
\eeq
For a general collection of free Weyl species, the equations of motion are
\beq
\d_{s_a}\psi^a=0
\quad\Rightarrow\quad
\psi^a=\text{function of }t-s_a x.
\eeq
For the Dirac fermion defined above, the classical Hamiltonian
density is
\beq
\scr{H}_\text{Dirac}
=-\sum_{a=1}^2s_a\psi^{a\dagger}\ii\d_1\psi^a.
\eeq
Note that the Dirac matrices are real, so $\psi^a$ and
$\psi^{a\dagger}$ transform in the same representation of the Lorentz group. This means that the gauge charges $q_a$ can be chosen nonnegative without loss of generality by conjugating individual Weyl species.

\clearpage

\phantomsection
\addcontentsline{toc}{section}{References}

\pdfbookmark[1]{References}{Refs}

\IfFileExists{current-draft-refs.bib}{%
  \bibliographystyle{utphys}%
  \bibliography{current-draft-refs}%
}{%
  \IfFileExists{Overleaf/current-draft-refs.bib}{%
    \bibliographystyle{utphys}%
    \bibliography{current-draft-refs}%
  }{%
    \section*{References}
    \PackageWarningNoLine{Draft}{Bibliography database current-draft-refs.bib
      is missing; citations will remain unresolved}%
  }%
}

\end{document}